\documentclass[12pt,a4paper]{article}

\RequirePackage[l2tabu, orthodox]{nag}

\usepackage{jheppub}
\usepackage{amsthm,amssymb,amsmath,epic,float}
\usepackage{rotating,epsfig,indentfirst,array,varioref}
\usepackage{appendix,tikz,mathtools}
\usepackage{setspace}
\usepackage{tikz}
\usetikzlibrary{decorations.pathreplacing}
\usetikzlibrary{calc}
\usetikzlibrary{positioning}
\usepackage{graphicx}  
\usepackage{bbold}

\let\emptyset\varnothing

\def\leq{\leqslant}
\def\geq{\geqslant}

\newdimen\tableauside\tableauside=1.0ex
\newdimen\tableaurule\tableaurule=0.4pt
\newdimen\tableaustep
\def\phantomhrule#1{\hbox{\vbox to0pt{\hrule height\tableaurule
width#1\vss}}}
\def\phantomvrule#1{\vbox{\hbox to0pt{\vrule width\tableaurule
height#1\hss}}}
\def\sqr{\vbox{%
  \phantomhrule\tableaustep

\hbox{\phantomvrule\tableaustep\kern\tableaustep\phantomvrule\tableaustep}%
  \hbox{\vbox{\phantomhrule\tableauside}\kern-\tableaurule}}}
\def\squares#1{\hbox{\count0=#1\noindent\loop\sqr
  \advance\count0 by-1 \ifnum\count0>0\repeat}}
\def\tableau#1{\vcenter{\offinterlineskip
  \tableaustep=\tableauside\advance\tableaustep by-\tableaurule
  \kern\normallineskip\hbox
    {\kern\normallineskip\vbox
      {\gettableau#1 0 }%
     \kern\normallineskip\kern\tableaurule}%
  \kern\normallineskip\kern\tableaurule}}
\def\gettableau#1 {\ifnum#1=0\let\next=\null\else
  \squares{#1}\let\next=\gettableau\fi\next}

\def\be{\begin{equation}}
\def\ee{\end{equation}}
\def\ba{\begin{array}}
\def\ea{\end{array}}

\restylefloat{figure}

\newcommand{\cA}{\mathcal{A}}

\newcommand{\cF}{\mathcal{F}}

\newcommand{\cH}{\mathcal{H}}

\newcommand{\cN}{\mathcal{N}}

\renewcommand{\mod}{\textup{mod}\,}

\usepackage{jheppub}
\makeatletter
\def\@fpheader{\vspace{-.1cm}}
\makeatother

\title{\boldmath AGT basis in SCFT for c=3/2 and Uglov Polynomials}

\author[a\,\dagger]{\;Vladimir\ Belavin,}
\author[b]{\;Abay\ Zhakenov}
\note[$\dagger$]{On leave from Lebedev Institute and the Institute for Information Transmission Problems, Moscow.}

\affiliation[a]{Physics Department, Ariel University, Ariel 40700, Israel}
\affiliation[b]{Moscow Institute of Physics and Technology, 141700 Dolgoprudny, Russia}
\emailAdd{vlbelavin@gmail.com}
\emailAdd{zhakenov.ak@phystech.edu}

\abstract{
AGT allows one to compute conformal blocks of d = 2 CFT for a large class of chiral CFT algebras.  This is related  to the existence of a certain orthogonal basis in the module of the (extended) chiral algebra. The elements of the basis are eigenvectors of a certain integrable model, labeled in general by N-tuples of Young diagrams. In particular, it was found that in the Virasoro case these vectors are expressed in terms of Jack polynomials, labeled by 2-tuples of ordinary Young diagrams, and for the super-Virasoro case they are related to Uglov polynomials, labeled by two colored Young diagrams. In the case of a generic central charge this statement was checked in the case when one of the Young diagrams is empty. In this note we study the N=1 SCFT and construct 4 point correlation function using the basis. To this end we need to clarify the connection between basis elements and Uglov polynomials, we also need to use two bosonizations and their connection to the reflection operator.  For the central charge $c=3/2$ we checked that there is a connection with the Uglov polynomials for the whole set of diagrams. 
}


\begin{document}
\maketitle
\flushbottom

   \label{intro}
    \section{Introduction}
    
The AGT correspondence~\cite{Alday:2009aq}  connects  two-dimensional conformal field theory with Virasoro symmetry and a class of four-dimensional $\mathcal{N}=2$ supersymmetric $SU(2)$ quiver gauge field theories. For recent reviews see, e.g., \cite{Rodger2013,Szabo:2015wua}. In particular, it connects the instanton part of Nekrasov partition functions \cite{Nekrasov:2002qd,Flume:2002az,Nekrasov:2003rj} to correlation functions of  Liouville theory. This relation is generalized to CFTs with additional symmetries, such as affine and $\mathcal{W}_k$-symmetry  in~\cite{Alday:2010vg,Wyllard:2009hg,Mironov:2009by,Belavin:2012qh},  with extended supersymmetry in~\cite{Belavin:2011pp, Bonelli:2011jx, Belavin:2011tb,Bonelli:2011kv,Ito:2011mw, Belavin:2012aa, Belavin:2012eg,Belavin:2012uf} and to the CFT with parafermion algebra in~\cite{Wyllard:2011mn,Alfimov:2011ju}. 
    
The connection between the instanton part of Nekrasov partition functions and CFT correlation functions is related to the existence of a special basis in the extended CFT, such that the matrix elements of the vertex operators in this basis are equal to the Nekrasov function $Z_{\text{bif}}$, which  will be given explicitly below.

In~\cite{Belavin:2011pp,Belavin:2011sw} it was suggested that conformal blocks in the CFT with the coset symmetry $\widehat{gl}_r(n)/\widehat{gl}_r(n-p)$
    correspond to the instanton partition functions on $\mathbb{C}^{2} / \mathbb{Z}_{p}$ in the $SU(r)$ gauge theory. This coset symmetry by the level-rang duality is 
    \be\label{Coset-A}
    \mathcal{A}(r, p) = \mathcal{H} \times \widehat{\mathfrak{s l}}(p)_{r} \times \frac{\widehat{\mathfrak{s l}}(r)_{p} \times \widehat{\mathfrak{s l}}(r)_{n-p}}{\widehat{\mathfrak{s l}}(r)_{n}}\;.
    \ee 
 Here $\cH$ is the Heisenberg algebra, $\widehat{\mathfrak{s l}}(r)_p$ is the affine Lie algebra and $n$ is related to the so-called equivariant parameter, which defines the central charge of the dual CFT. The original AGT correspondence arises for $r=2$ and $p=1$. The coset~\eqref{Coset-A} defines the extension of the original version to the three-parametric family of CFTs with the parameters $r,p$ and $n$. With respect to the dependence on $n$, two possibilities are distinguished: a CFT with continuous operator product expansions (OPE) like the Liouville theory and a rational CFT like minimal models. For the specific features of the AGT correspondence in the later case see, e.g.~\cite{Bershtein:2014qma,Alkalaev:2014sma,Belavin:2015ria,Manabe:2020etw}.

    Up to now the special basis was constructed for the  Virasoro case in~\cite{Alba:2010qc} and for the case $r>2$ and $p=1$ in~\cite{Fateev:2011hq,Albion:2020qhl}.
   The case $r=2$, $p=2$, related to $\cN=1$ SCFT, was considered in~\cite{Belavin:2012eg}. It was  claimed there that the basis in the free field representation of the algebra $\mathcal{A}(2,2)$ can be expressed through the so-called Uglov polynomials \cite{Uglov:1997ia,uglov1997symmetric}, which are obtained in the limit $q,~ t \rightarrow -1$ from Macdonald polynomials. The use of the connection with the Uglov polynomials in this context requires
   introducing two bosonizations for the representation of $\cA(2,2)$.
   The first one is defined in terms of free fields generators by means of the Feigin-Fuchs bosonization of the super-Virasoro algebra and Fateev-Zamolodchikov bosonization of $\widehat{\mathfrak{sl}}(2)_2$. The second one requires the implementation of two additional representations of the Heisenberg algebra, with the generators $a^{(1,2)}_k$. The connection between the generators $a^{(1,2)}_k$ and the generators of $\cA(2,2)$ plays an important role in the construction of the basis elements.

In this paper we are dealing with the case $r=2$, $p=2$ and consider the construction of the correlation functions in terms of elements of the special basis.  
In the work \cite{Belavin:2011tb} the problem of constructing correlation functions from AGT was considered. The expressions were derived using OPE, where the coefficients for each level of the descendants' contribution was written in terms of the instanton partition functions.
On the other hand, in work \cite{Belavin:2012eg} the problem of constructing of the AGT basis was studied. In this note we analyze the connection between these two problems. We consider the 4-point function on the sphere. We find that the relation between the two $\cA(2,2)$ bosonizations, when applied to the first levels in the OPE expansion, requires a more detailed consideration. In particular, to construct the special basis in terms of the Uglov polynomials on the first levels, we need to express the components of the polynomials (power-sum polynomials) in terms of $\cA(2,2)$ algebra elements, which requires a particular formulation of their relation. 
In order to represent the OPE coefficients, that define the conformal block, in terms of the special basis we also have to formulate the conjugation rule for the basis elements. In addition, the construction of a general basis element requires two Feigin-Fuchs bosonizations connected by means of a certain reflection transformation which we also analyze. 
	
The special basis is labeled by pairs of two-colored diagrams (left and right), where the left diagrams are associated with one set of generators and the right diagrams with another set. These sets are connected by the reflection transformation. For the generic value of the central charge $c$, because of the noncommutativity of the sets, the basis elements are known only for a subclass labeled by pairs with one empty diagram (left or right). For the particular case $c=3/2$, the left and right sets commute, and each element of the special basis is given by a product of the Uglov polynomials corresponding to each diagram. It is possible in this case to express OPE coefficients in terms of the basis elements and to get an expression for the correlation function in a closed form.

The plan of the paper is the following. In Section~\ref{AGTforN1} we discuss the free field representation of the $\mathcal{A}(2,2)$ algebra, the AGT basis and recall the construction of 4-point conformal blocks in terms of the so-called chain vectors. In Section~\ref{bosonization} we obtain the elements of the special basis in terms of the Uglov polynomials using two bosonizations of $\cA(2,2)$. In Section~\ref{c1} we consider in detail the $c=3/2$ case and verify the scalar products of the  basis with the chain vectors against the AGT results. In Section~\ref{concl} we present our conclusion. Important properties of the Macdonald and Uglov polynomials are collected in Appendix~\ref{UglovPol}, explicit expressions for the elements of the special basis at the first four levels are written in Appendix~\ref{SpecialBas}, Appendix~\ref{ChainVectorsAndSpecBasis} consists of some formulas related to Nekrasov's functions and chain vectors, Appendix~\ref{Level2} contains detailed computations of the scalar products of the basis elements with the chain vector on the level 2.

\section{AGT and SVir CFT}
\label{AGTforN1}

\paragraph{Chiral symmetry algebra.}
The algebra $\mathcal{A}(2,2)$, eq.~\eqref{Coset-A}, is the tensor product of NSR (super-Virasoro), $\widehat{\text{sl}}(2)_2$ and Heisenberg algebras, $\mathcal{A}(2,2)=\cH\oplus\widehat{\text{sl}}(2)_2 \oplus \text{NSR}$. To construct the special basis it will be essential to represent subalgebras in terms of free fields. For $\cH$ and $\widehat{\text{sl}}(2)_2$ it is formulated below, while for NSR algebra the bosonization is discussed in Section~\ref{bosonization}.
The NSR algebra commutation relations read
\begin{equation}\label{OperCom}
\begin{aligned} 
\left[L_{n}, L_{m}\right] &=(n-m) L_{n+m}+\frac{\hat{c}}{8}\left(n^{3}-n\right) \delta_{n+m}\;, \\ 
\left\{G_{r}, G_{s}\right\} &=2 L_{r+s}+\frac{1}{2} \hat{c}\left(r^{2}-\frac{1}{4}\right) \delta_{r+s}\;, \\
\left[L_{n}, G_{r}\right] &=\left(\frac{1}{2} n-r\right) G_{n+r}\;, 
\end{aligned}
\end{equation}
where $n,m$ are integer and $r,s$ are either integer (R sector), or half-integer (NS sector). In what follows we focus on the NS sector.  
We use the following parametrization of the central charge
\begin{equation}
\hat{c}=1+2Q^2\;,~~~~~~Q=b+\frac{1}{b}
\end{equation}
and $c\equiv\frac{3\hat{c}}2$. In particular, $\hat{c}=1$ corresponds to the free theory.  
\newline
The affine Lie Algebra of the level $k$, $\widehat{\mathrm{sl}}(2)_k$, is defined by the commutation relations
\be
\ba{l}
\left[e_{n}, e_{m}\right]=\left[f_{n}, f_{m}\right]=0\;, \quad\left[e_{n}, f_{m}\right]=h_{n+m}+n \delta_{n+m} k\;,\\
\left[h_{n}, e_{m}\right]=2 e_{n+m}\;, \quad\left[h_{n}, f_{m}\right]=-2 f_{n+m}\;, \quad\left[h_{n}, h_{m}\right]=2 n \delta_{n+m} k\;.
\ea
\ee
Its integrable representation $\mathcal{L}_{h, k}$ with the highest vector $|v\rangle$ is defined by
\be
e_{n} v=0\;,\quad \text{for } n \geq 0 \;; \quad f_{n} v=h_{n} |v\rangle=0\;,\quad \text {for } n>0 \;; \quad h_{0} |v\rangle=h |v\rangle\;.
\ee
We use Fateev-Zamolodchikov realization \cite{Fateev:1985mm} to give the free field representation of $\widehat{\mathrm{sl}}(2)_2$  
\be
\ba{l}\label{FateevZamolodchikov}
\left[h_{n}, h_{m}\right]=4 n \delta_{n+m, 0}\;, \quad\left\{\chi_{r}, \chi_{s}\right\}=\delta_{r+s, 0}\;, \quad\left[h_{n}, \chi_{r}\right]=0\;,\\ 
\left[D, h_{n}\right]=0\;, \quad \text {for } n \neq 0\;, \quad\left[h_{0}, D\right]=D\;, \quad\left[\chi_{r}, D\right]=0\;,
\ea
\ee
where  $h_n$ are the generators of the Heisenberg subalgebra $\mathcal{H}^{\mathrm{sl}(2)}$ and $\chi_r$ are the generators of the Majorana Fermion subalgebra $\cF^{\mathrm{sl}(2)}$. We denote the Fock representation of the algebra $\cH^{\mathrm{sl}(2)} \oplus \cF^{\mathrm{sl}(2)}$ with the vacuum vector $|v_J\rangle$ by $\mathrm{F}_{J}$:
\be\label{v_J}
h_{n} |v_{J}\rangle=0\;,\quad \chi_{r} |v_{J}\rangle=0 \quad \text {for } n, r>0\;; \quad h_{0} |v_{J}\rangle=J |v_{J}\rangle\;.
\ee
From \eqref{FateevZamolodchikov} it follows that the operator $D$ maps from one Fock space to another 
\be\label{D}
D:\mathrm{F}_{J} \rightarrow \mathrm{F}_{J+1}
\ee
and hence just shifts the value of $h_0$, for convenience we will work with $h_0=0$.
The generators $w_n$ of the Heisenberg algebra $\mathcal{H}$ are defined by the commutation relation
\be\label{CommutationRelationsW}
\left[w_{n}, w_{m}\right]=4 n \delta_{n+m,0}\;.
\ee
The representation of this algebra with the highest vector $\tilde{|v\rangle}$ is defined by
\be\label{vW}
w_{n}\tilde{|v\rangle}=0\;,\quad\text{for }n>0\;.
\ee
The described representation of the algebra $\cH\oplus\cH^{\mathrm{sl}(2)}\oplus\cF^{\mathrm{sl}(2)}\oplus\mathrm{NS}$ gives the representation of $\mathcal{A}(2,2)$ for each $h_0$.

\paragraph{Vertex operators and correlation functions.} For the primary NS field $\Phi_\Delta$ we use the following parametrization of the conformal dimension $\Delta(\alpha)=\frac{1}{2}\alpha(Q-\alpha)$, the complex parameter $\alpha$ is expressed in terms of the so-called momentum $P$, as $\alpha=Q/2+P$, that is $\Delta(P)=\frac{1}{2}\left(Q^2/4-P^2\right)$. The upper component of the primary super doublet is $ \Psi_\Delta=G_{-1/2} \Phi_\Delta$ with the conformal dimension $\Delta+1/2$.

We shall consider the 4-point conformal block on the sphere.
The $s$-channel expansion for the lower components is
\begin{align}
\langle\Phi_1(q)\Phi_2(0)\Phi_3(1)\Phi_4(\infty)\rangle=
\sum_{\Delta}\bigg[&C_{12}^{\Delta}C_{34}^{\Delta}
F_0(\Delta,\Delta_i,c,q)F_0(\Delta,\Delta_i,c,\bar q)
\nonumber\\
&{}+\tilde C_{12}^{\Delta}\tilde C_{34}^{\Delta}
F_1(\Delta,\Delta_i,c,q) F_1(\Delta,\Delta_i,c,\bar q)\bigg]\;,
\label{corrfun}
\end{align}
where $q$ is the 4-point harmonic ratio, $C_{ij}^k$ and $\tilde C_{ij}^k$ are structure constants of the operator algebra, and 
\be
\begin{aligned}\label{ConfBlockDef}
&F_0(\Delta,\Delta_i,c,q)=q^{\Delta-\Delta_1-\Delta_2}
\sum^{N\text{ integer}}_{N\ge0}q^N~{}_{12}^{SV}\langle N|N\rangle_{34}^{SV}\;,
\\
&F_1(\Delta,\Delta_i,c,q)=q^{\Delta-\Delta_1-\Delta_2}
\sum^{N\text{ half-integer}}_{N>0}q^N~{}_{12}^{SV}\langle N|N\rangle^{SV}_{34}
\end{aligned}
\ee
are 4-point super conformal blocks~\cite{Belavin:2006zr}.  Here $|N\rangle^{SV}$ (the so-called chain vector) is the $N$th-level descendant contribution of
the intermediate state with the conformal dimension $\Delta$ arrising in the
operator product expansion $\Phi_1(q)\Phi_2(0)$:
\begin{equation}
[\Phi_1(q)\Phi_2(0)]_{\Delta}=
q^{\Delta-\Delta_1-\Delta_2}\sum_{N=0}^{\infty}q^N |N\rangle_{12}^{SV}\;.
\end{equation}
The vectors $|N\rangle_{12}^{SV}$ from eq.~\eqref{ConfBlockDef} and the vectors
$\widetilde{|N\rangle}{}_{12}^{SV}$ arising in
the OPE $\Psi_1(q)\Phi_2(0)$,
\begin{equation}
[\Psi_1(q)\Phi_2(0)]_{\Delta}=q^{\Delta-\Delta_1-\Delta_2-\frac{1}{2}}
\sum_{N=0}^{\infty} q^N \widetilde{|N\rangle}{}_{12}^{SV}\;,
\end{equation}
are the subject of the following recursion relations~\cite{Belavin:2006zr} with the initial condition $\langle 0 | 0\rangle=1$ 
\begin{equation}
\begin{cases}
G_k|N\rangle_{12}^{SV}={\widetilde{|N-k \rangle}}{}_{12}^{SV}\;,\\
G_k{\widetilde{|N\rangle}}{}_{12}^{SV}=
[\Delta+2k\Delta_1-\Delta_2+N-k]|N-k\rangle{}_{12}^{SV}\;,\end{cases}
\label{chain}
\end{equation}
where $k>0$, which allow one to fix these vectors level by level. 

The AGT provides the explicit expressions for the conformal blocks. In this context we are dealing with the  ``dressed'' primary fields $V_\alpha$  defined as
\begin{equation}
V_\alpha=\mathcal{V}_\alpha \cdot \Phi_\alpha\;,
\end{equation}
where $\mathcal{V}_\alpha$ acts in $\cH\oplus\widehat{\text{sl}}(2)_2$ sector and explicitly is 
\begin{equation}
\mathcal{V}_{\alpha}(z)=\exp \left(i(\alpha-Q) \sum_{n=1}^{\infty} \frac{w_{-n} z^{n}}{-2 n}\right) \exp \left(i \alpha \sum_{n=1}^{\infty} \frac{w_{n} z^{-n}}{2 n}\right)\;.
\end{equation}
The commutation relations of the field  $\mathcal{V}_{\alpha}(z)$ with the generators $h_n,~\chi_s$ and $w_m$ can be obtained from~\eqref{FateevZamolodchikov} and \eqref{CommutationRelationsW}: 
\be\label{commRel}
\ba{l}
\left[w_n, \mathcal{V}_\alpha(z)\right]=2i\left(Q-\alpha\right)z^n\mathcal{V}_\alpha(z),\quad n>0\;,\\
\left[w_n, \mathcal{V}_\alpha(z)\right]=-2i\alpha z^{n}\mathcal{V}_\alpha(z),\quad n<0\;, \\
\left[h_n, \mathcal{V}_\alpha(z)\right]=\left\{\chi_r, \mathcal{V}_\alpha(z)\right\}=0\;.
\ea
\ee
To formulate the AGT relation for the 4-point blocks we introduce functions 
\begin{align}
&B_ i(q) = G_i(q)F_i(q)\;\label{ConfBlockBFGDef},
\end{align}
where $F_{0,1}(q)$ are defined in \eqref{ConfBlockDef} and $G_{0,1}(q)$ stand for $\cH\oplus\widehat{\text{sl}}(2)_2$ conformal blocks
\be
\begin{aligned}\label{ConfBlockGDef}
&G_0(q)=(1-q)^{(\frac{Q}2+P_1)(\frac{Q}2-P_3)}\;,\\
&G_1(q)=\frac{1}{2}(1-q)^{(\frac{Q}2+P_1)(\frac{Q}2-P_3)}\;.
\end{aligned}
\ee
For our purposes it is instructive to interpret these equations in terms of the chain vectors, $|N\rangle^{H\oplus\widehat{\text{sl}}(2)_2}$ which define Nth level descendant contribution\footnote{The grading in the module of $\cH\oplus\widehat{\text{sl}}(2)_2$ is the minus sum of the generator indices.}   to the OPE  $\mathcal{V}_1(z)\mathcal{V}_2(0)$. These chain vectors are defined by 
\be
\ba{l}\label{HSl22ChainCommRel}
w_n |N,\alpha_3, \alpha_4\rangle^{H\oplus\widehat{\text{sl}}(2)_2}=2i(Q-\alpha_3)|N-n,\alpha_1, \alpha_2\rangle^{H\oplus\widehat{\text{sl}}(2)_2}\;,\\[0.5em]
^{H\oplus\widehat{\text{sl}}(2)_2}\langle N,\alpha_1, \alpha_2|w_{-n}=~^{H\oplus\widehat{\text{sl}}(2)_2}\langle N-n,\alpha_1, \alpha_2|2i\alpha_1\;,
\ea
\ee
and $h_{n>0},\psi_{r>0}$ annihilate them. The highest vector $|0\rangle^{H\oplus\widehat{\text{sl}}(2)_2}=|v_J\rangle \otimes \tilde{|v\rangle}$. Explicitly one can write
\be
\ba{l}
|N,\alpha_3, \alpha_4\rangle^{H\oplus\widehat{\text{sl}}(2)_2}=\sum_{r_1,r_2,...}\prod_{l=1}^\infty \frac{C^{r_l}}{r_l! l^{r_l}} |r_1,r_2,...\rangle\;,\\[0.5em]
^{H\oplus\widehat{\text{sl}}(2)_2}\langle N,\alpha_1, \alpha_2|=\sum_{k_1,k_2,...}\prod_{l=1}^\infty  \langle k_1,k_2,\dots| \frac{E^{k_l}}{k_l! l^{k_l}}
\ea
\ee
with constants $E=\frac{i\alpha_1}{2},~C=\frac{i(Q-\alpha_3)}{2}$, and states $\langle k_1,k_2,\dots|=\langle 0|\dots w_2^{k_2}w_1^{k_1}$ and $|r_1,r_2,...\rangle= w_{-1}^{r_1}w_{-2}^{r_2}....|0\rangle$ with $k_1+k_2+\dots=N$ and $r_1+r_2+\dots=N$. Hence, the conformal blocks $G_i(q)$ from eqs.~\eqref{ConfBlockGDef} are nothing but
\be
 \begin{aligned}
&G_0(q)=\sum_N {}^{H\oplus\widehat{\text{sl}}(2)_2}\!\langle N,\alpha_1, \alpha_2|N,\alpha_3, \alpha_4\rangle^{H\oplus\widehat{\text{sl}}(2)_2} q^N\;,\\
&G_1(q)=\frac{1}{2}\sum_N {}^{H\oplus\widehat{\text{sl}}(2)_2}\!\langle N,\alpha_1, \alpha_2|N,\alpha_3, \alpha_4\rangle^{H\oplus\widehat{\text{sl}}(2)_2} q^N\;.
\end{aligned}
\ee
Here the factor $\frac12$ is due to the conventional normalization of the super conformal blocks, for details see~\cite{Belavin:2011tb}. 
Using \eqref{ConfBlockBFGDef}, the total conformal blocks can be written as 
\be
\begin{aligned}\label{ConfBlockBDef}
&B_0=q^{\Delta-\Delta_1-\Delta_2}
\sum^{N\text{ integer}}_{N\ge0}q^N{}_{12}\langle N|N\rangle_{34}\;,
\\
&B_1=\frac{1}{2}~q^{\Delta-\Delta_1-\Delta_2}
\sum^{N\text{ half-integer}}_{N>0}q^N{}_{12}\langle N|N\rangle_{34}\;,
\end{aligned}
\ee
where 
\be\label{ChainVector}
|N\rangle_{ij}=\sum_{N_1+N_2=N}|N_1\rangle_{ij}^{SV} |N_2\rangle_{ij}^{H\oplus\widehat{\text{sl}}(2)_2}
\ee
for both integer and half-integer $N$.

From AGT it follows (more details in \cite{Belavin:2011tb}) that
\be\label{ChainNorms}
\ba{l}
{}_{12}\langle N|N\rangle_{34}=\sum_{\vec{\lambda}^0, N_{+}(\vec{\lambda}^0)=N \atop N_{-}(\vec{\lambda}^0)=N} Z_{\mathrm{f}}^{\mathrm{sym}}\left(\mu_{i}, \vec{P}, \vec{\lambda}^0\right) Z_{\mathrm{vec}}^{\mathrm{sym}}(\vec{P}, \vec{\lambda}^0)\;,\quad\text{for integer }N\;,\\
	{}_{12}\langle N|N\rangle_{34}=2\sum_{\vec{\lambda}^1, N_{+}(\vec{\lambda}^1)=N-\frac{1}{2} \atop N_{-}(\bar{\lambda}^1)=N+\frac{1}{2}} Z_{\mathrm{f}}^{\mathrm{sym}}\left(\mu_{i}, \vec{P}, \vec{\lambda}^1\right) Z_{\mathrm{vec}}^{\mathrm{sym}}(\vec{P}, \vec{\lambda}^1)\;,\quad\text{for half-integer }N\;,
\ea
\ee
where $\vec{\lambda}^\sigma$ is a pair of two-colored (in chess coloring) Young diagrams $\lambda_1^\sigma$ and $\lambda_2^\sigma$ with the color of the angle cell $\sigma$ being labeled by $0$ (white) or $1$ (black). The sums in \eqref{ChainNorms} go over pairs of diagrams with $N_{+}(\vec{\lambda}^\sigma)=N_{+}(\lambda_1^\sigma)+N_{+}(\lambda_2^\sigma)$ white colored and $N_{-}(\vec{\lambda}^\sigma)=$ $=N_{-}(\lambda_1^\sigma)+N_{-}(\lambda_2^\sigma)$ black colored cells\footnote{For convenience we change the color of the angle cell for half-integer values of $N$ from white (as in \cite{Belavin:2011tb}) to black.}. The $Z_{\mathrm{f}}^{\mathrm{sym}}$ and $Z_{\text {vec }}^{\text {sym }}$ are just the standard AGT functions, which are given in Appendix~\ref{ChainVectorsAndSpecBasis} . Parameters of these functions are related with the parameters of the conformal block as follows
\be
\ba{l}\label{mu}
\mu_{1}=\frac{Q}{2}+\left(P_{1}+P_{2}\right)\;, \quad \mu_{2}=\frac{Q}{2}+\left(P_{1}-P_{2}\right)\;,\\
\mu_{3}=\frac{Q}{2}-\left(P_{3}+P_{4}\right)\;, \quad \mu_{4}=\frac{Q}{2}-\left(P_{3}-P_{4}\right)\;,\\
\ea
\ee
and
\be
\vec{P}=(P,-P)\;.
\ee

\paragraph{Special basis of states in CFT with A(2,2) symmetry.}
It follows from
AGT that there exists orthogonal basis  $\left|P\right\rangle_{\vec{\lambda}^\sigma}$ in the representation space of $\cA(2,2)$, labeled by a pair of Young diagrams $\vec{\lambda^\sigma}=(\lambda_1^\sigma, \lambda_2^{\sigma})$ such that
\begin{equation}\label{AGTrel}
\frac{_{\vec{\mu}^\sigma}\left\langle P^{\prime}\left|\Phi_{\alpha}\right| P\right\rangle_{\vec{\lambda}^{\tilde{\sigma}}}}{\left\langle P^{\prime}\left|\Phi_{\alpha}\right| P\right\rangle}= Z_{\text{bif}}\left(\alpha\left|P',\vec{\mu}^\sigma, P, \vec{\lambda}^{\tilde{\sigma}}\right.\right)\;,
\end{equation}
where $\left| P \right\rangle$ is an NS primary state with the conformal dimension $\Delta(P)=\frac{1}{2}(Q^2/4-P^2)$. The function $Z_{\text{bif}}$ is given in Appendix~\ref{ChainVectorsAndSpecBasis}.
This special basis respects the grading:
\be\label{Pbasis}
h_0|P\rangle_{\vec{\lambda}^\sigma}=\left(2d(\vec{\lambda}^\sigma)+2\sigma\right)|P\rangle_{\vec{\lambda}^\sigma}\;,\qquad
L_0|P\rangle_{\vec{\lambda}^\sigma}=\left(\frac{2|\vec{\lambda}^\sigma|+h_0}{4}+\Delta\right)|P\rangle_{\vec{\lambda}^\sigma}\;,
\ee
with $d(\lambda^\sigma)=N_0(\lambda^\sigma)-N_1(\lambda^\sigma)$ being the diference between the number of black and white cells in diagram $\lambda^\sigma$ and $d(\vec{\lambda}^\sigma)=d(\lambda_1^\sigma)+d(\lambda_2^\sigma)$ --- the same for the pair of diagrams. The choice of $h_0\in\mathbb{Z}$ is arbitrary but for convenience we take $h_0=0$ since other values $h_0$ can be obtained acting by the operator $D$, eq.~\eqref{D}. Since $h_0=0$ the color $\sigma$ of the angle cells for the pair of diagrams $\vec{\lambda}^\sigma$ is defined by the total number of cells $|\vec{\lambda}^\sigma|$:
\be\label{sigma}
\ba{l}
\sigma=0 \leftrightarrow |\vec{\lambda}^\sigma|\text{ is even}\;,\\
\sigma=1 \leftrightarrow |\vec{\lambda}^\sigma|\text{ is odd}\;.
\ea
\ee 
Let us consider the first few representatives of the basis \cite{Belavin:2012eg}. 
\newline
At the level $L_0=\Delta$ we have the ground state $|P\rangle$.\newline
At the level $L_0=\Delta+\frac{1}{2}$ there are two states:
\begin{equation}
\begin{array}{l}
{|P\rangle_{\left((1)^{1}, \varnothing^{1}\right)}=-\sqrt{2}\left(G_{-1 / 2}+i \frac{Q+2 P}{2} \chi_{-1 / 2}\right)|P\rangle\;,}\\
{|P\rangle_{\left(\varnothing^{1},(1)^{1}\right)}=-\sqrt{2}\left(G_{-1 / 2}+i \frac{Q-2 P}{2} \chi_{-1 / 2}\right)|P\rangle}\;.
\end{array}
\end{equation}
Four states for the level $L_0=\Delta+1$ are:
\begin{equation}\label{basis1}
\begin{array}{l}
{|P\rangle_{\left((2)^{0}, \varnothing^{0}\right)}=\left(-2 L_{-1}-\frac{2 i}{b} \chi_{-1 / 2} G_{-1 / 2}-\frac{i(Q+2 P)}{2} w_{-1}-\frac{Q+2 P}{2 b} h_{-1}\right)|P\rangle}\;,\\
{ |P\rangle_{\left((1,1)^{0}, \varnothing^0\right)} =\left(-2 L_{-1}-2 b i \chi_{-1 / 2} G_{-1 / 2}-\frac{i(Q+2 P)}{2} w_{-1}-\frac{b(Q+2 P)}{2} h_{-1}\right)|P\rangle\;,} \\ 
{|P\rangle_{\left(\varnothing^{0},(2)^{0}\right)} =\left(-2 L_{-1}-\frac{2 i}{b} \chi_{-1 / 2} G_{-1 / 2}-\frac{i(Q-2 P)}{2} w_{-1}-\frac{Q-2 P}{2 b} h_{-1}\right)|P\rangle\;, }\\
{|P\rangle_{\left(\varnothing^0,(1,1)^0\right)}=\left(-2 L_{-1}-2 b i \chi_{-1 / 2} G_{-1 / 2}-\frac{i(Q-2 P)}{2} w_{-1}-\frac{b(Q-2 P)}{2} h_{-1}\right)|P\rangle\;.}
\end{array}
\end{equation}
The basis we are going to use in this paper is denoted as $|\vec{\lambda}\rangle=|\lambda_1,\lambda_2\rangle$. Its connection with the basis $|P\rangle_{\vec{\lambda}^\sigma}$, eq.~\eqref{Pbasis}, is the following
\be\label{OldNewOrthBasis}
|P\rangle_{\lambda_1^\sigma,\lambda_2^\sigma}=\Omega_{\vec{\lambda}}(P) |\lambda_1,\lambda_2\rangle\;.\\
\ee
We recall that the $\sigma$ parameter is not free, see eq.~\eqref{sigma}, and that is why is does not appear on the right hand side.
 This equation is the definition of the basis $|\vec{\lambda}\rangle$. The normalization function $\Omega_{\vec{\lambda}}(P)$ is known \cite{Belavin:2012eg} only for the pairs with one empty diagram:
\be
\ba{l}
\Omega_{\lambda,\varnothing}(P)=\prod_{s \in \lambda, i+j \equiv 0 \bmod 2}\left(2 P+i b+j b^{-1}\right)\;,\\
\Omega_{\varnothing,\lambda}(P)=\prod_{s \in \lambda, i+j \equiv 0 \bmod 2}\left(-2 P+i b+j b^{-1}\right)\;,
\ea
\ee
where $i$ and $j$ are coordinates of a cell in vertical and horizontal axes correspondingly. One can notice that $\Omega_{\lambda,\varnothing}(P)=\Omega_{\varnothing,\lambda}(-P)$.
The basis $|\vec{\lambda}\rangle$ from \eqref{OldNewOrthBasis} is convenient because as we will see after bosonization its coefficients do not depend on a momentum $P$.

\paragraph{Conjugation rule.}

The conjugation rule is not uniquely defined. This non-uniqueness manifests itself in the concrete form of the scalar products matrix\footnote{In \cite{Alba:2010qc} was used another conjugation rule where $P$ does not change  sign and the order of the diagrams remains the same. This conjugation leads to the antidiagonal matrix of the scalar products.}. We work with the diagonal one. The corresponding conjugation rule \cite{Belavin:2011js} is the following: 
for any complex function $f(P)$, $\left(f(P)|P\rangle\right)^+=\langle P|f(-P)$ and $\left(|\lambda_2,\lambda_1\rangle\right)^+ = \langle \lambda_1,\lambda_2|$.

For example:
\be
_{\lambda_1^\sigma,\lambda_2^\sigma}\langle P|=\left(|P\rangle_{\lambda_2^\sigma,\lambda_1^\sigma}\right)^+=\left(\Omega_{\lambda_2,\lambda_1}(P) |\lambda_2,\lambda_1\rangle\right)^+ =\Omega_{\lambda_2,\lambda_1}(-P)  \langle \lambda_1,\lambda_2|\;.
\ee 

\paragraph{Scalar products of the chain vectors with the basis elements.}
Using orthogonality of the special basis and eq.~\eqref{ChainNorms} we get
\be\label{ChainSpecBasis1}
\langle \vec{\lambda}|N\rangle_{34}=
\begin{cases}
	\frac{\prod_{\alpha=3}^{4} \prod_{s \in \lambda_{\alpha},s-\mathrm{white}}\left(\phi\left(P_{\alpha}, s\right)+\mu_{i}\right)\left(\phi\left(P_{\alpha}, s\right)+\mu_{j}\right)}{\Omega_{(\lambda_1,\lambda_2)}(P)}\;,\quad\text{for integer }N\;,\\
	\frac{-\sqrt{2}\prod_{\alpha=3}^{4} \prod_{s \in \lambda_{\alpha},s-\mathrm{white}}\left(\phi\left(P_{\alpha}, s\right)+\mu_{i}\right)\left(\phi\left(P_{\alpha}, s\right)+\mu_{j}\right)}{\Omega_{(\lambda_1,\lambda_2)}(P)}\;,\quad\text{for half-integer }N\;,
\end{cases}
\ee
\be\label{ChainSpecBasis2}
_{12}\langle N | \vec{\lambda} \rangle = 
\begin{cases}
	\frac{\prod_{\alpha=1}^{2} \prod_{s \in \lambda_{\alpha},s-\mathrm{white}}\left(\phi\left(P_{\alpha}, s\right)+\mu_{i}\right)\left(\phi\left(P_{\alpha}, s\right)+\mu_{j}\right)}{\Omega_{(\lambda_1,\lambda_2)}(P)}\;,\quad\text{for integer }N\\
	\frac{-\sqrt{2}\prod_{\alpha=1}^{2} \prod_{s \in \lambda_{\alpha},s-\mathrm{white}}\left(\phi\left(P_{\alpha}, s\right)+\mu_{i}\right)\left(\phi\left(P_{\alpha}, s\right)+\mu_{j}\right)}{\Omega_{(\lambda_1,\lambda_2)}(P)}\;,\quad\text{for half-integer }N\;.
\end{cases}
\ee
Details can be found in Appendix~\ref{ChainVectorsAndSpecBasis}.

\section{Two bosonizations and super-Liouville reflection operator}
\label{bosonization}

In the previous section we introduced the AGT basis. It appears that the subclass of the basis of the form $|\lambda,\varnothing\rangle$ or $|\varnothing,\lambda\rangle$ 
can be expressed in terms of the Uglov polynomials. To this end one has to use two Feigin-Fuchs bosonizations of the NS algebra. In this section we describe the bosonizations and introduce the ingredients for constructing the elements of the basis for this subclass.
\paragraph{Two bosonizations of NS algebra.}
We consider the algebra 
\be\label{UEAlgebra}
\left[c_{n}, c_{m}\right]=n \delta_{n+m, 0}\;, \quad \quad\left\{\psi_{r}\;, \psi_{s}\right\}=\delta_{r+s, 0}\;, \quad \left[x,y\right]=0\;,
\ee
with $c_n$, $n\in\mathbb{Z}\backslash \{ 0\}$ being boson generators, $\psi_r$, $r\in\mathbb{Z}+\frac{1}{2}$ -- fermion generators, $x,y\in(c_n,\psi_r,\hat{P})$ and $x\neq y$. The NS algebra is embedded into the universal enveloping algebra of~\eqref{UEAlgebra} as
\begin{equation}
\begin{aligned}\label{NsBos}
&L_{n}=\frac{1}{2} \sum_{k \neq 0, n} c_{k} c_{n-k}+\frac{1}{2} \sum_{r}\left(r-\frac{n}{2}\right) \psi_{n-r} \psi_{r}+\frac{i}{2}(Q n - 2 \hat{P}) c_{n}\;,\\
&L_{0}=\sum_{k>0} c_{-k} c_{k}+\sum_{r>0} r \psi_{-r} \psi_{r}+\frac{1}{2}\left(\frac{Q^{2}}{4}-\hat{P}^{2}\right)\;,\\
&G_{r}=\sum_{n \neq 0} c_{n} \psi_{r-n}+i(Q r - \hat{P}) \psi_{r}\;.
\end{aligned}
\end{equation}
The  highest-weight representation of the super-Virasoro algebra is the Fock space with the vacuum vector $|P\rangle$ such that $\hat{P} |P\rangle= P |P\rangle$ and
\be
\ba{l}
c_k|P\rangle=0\;,\quad k>0\;,\qquad \psi_r |P\rangle = 0\;, \quad r>0\;.
\ea
\ee
The commutation relations of the generators $c_k,\psi_r$ with the NS generators are
\be
\ba{l}
\left[c_k,L_n\right]=kc_{k+n} (1-\delta_{k+n,0}) - \dfrac{ik}{2}(kQ+2P) \delta_{k+n,0}\;,\\
\left[\psi_r,L_n\right]=\left(r+\frac{n}{2}\right)\psi_{n+r}\;,\quad \left[c_k,G_l\right]=k\psi_{k+l}\;,\\
\left\{\psi_r,G_l\right\}=c_{r+l} (1-\delta_{r+l,0}) - \dfrac{i}{2}(2rQ+2P)\delta_{r+l,0}\;.
\ea
\ee
In the universal enveloping algebra \eqref{UEAlgebra} one can introduce the second set of generators, $c_k^\mathrm{R}, \psi_r^\mathrm{R}$. They are related to another possible Feigin-Fuchs representation which differs from \eqref{NsBos} by the sign of $\hat{P}$ :
\begin{equation}
\begin{aligned}\label{RNsBos}
&L_{n}=\frac{1}{2} \sum_{k \neq 0, n} c^\mathrm{R}_{k} c^\mathrm{R}_{n-k}+\frac{1}{2} \sum_{r}\left(r-\frac{n}{2}\right) \psi^\mathrm{R}_{n-r} \psi^\mathrm{R}_{r}+\frac{i}{2}(Q n + 2 \hat{P}) c^\mathrm{R}_{n}\;,\\
&G_{r}=\sum_{n \neq 0} c^\mathrm{R}_{n} \psi^\mathrm{R}_{r-n}+i(Q r + \hat{P}) \psi^\mathrm{R}_{r}\;.
\end{aligned}
\end{equation}
The sets $c_k, \psi_r$ and $c_k^\mathrm{R}, \psi_r^\mathrm{R}$ are connected by some nonlinear, so-called reflection transformation which is unknown in the closed form. Nevertheless, the first terms of the formal expansion can be found explicitly by considering the states on the lower levels. For instance, solving the following system 
\be
\ba{c}
L_{\lambda}G_\mu(c_k^\mathrm{R},\psi_r^\mathrm{R},-\hat{P})|P\rangle = L_{\lambda}G_\mu(c_k,\psi_r,\hat{P})|P\rangle\;,
\ea
\ee
for the states on the levels from 1/2 to 2 for the generators $\psi_{-1/2}^\mathrm{R}\;$, $c_{-1}^\mathrm{R}\;$, $\psi_{-3/2}^\mathrm{R}$ and $c_{-2}^\mathrm{R}$ one finds
\be
\begin{aligned}\label{ReflectedOper1}
&\psi_{-1/2}^\mathrm{R}=\frac{Q+2P}{Q-2P}\psi_{-1/2}+\dots\;,\\
&c_{-1}^\mathrm{R}=\frac{Q+2P}{Q-2P}c_{-1} +\frac{32 P Q}{(2 P-Q) (2 P+Q) \left(4 P^2-8 P Q+3 Q^2+4\right)}c_{-1}\psi_{-1/2}\psi_{1/2}+\\
&+\frac{16 i P Q}{8 P^3-12 P^2 Q-2 P Q^2+8 P+3 Q^3+4 Q}\psi_{-3/2}\psi_{1/2}+\\
&+\frac{4 i \left(2 P^2 Q-3 P Q^2\right)}{(2 P+Q) \left(4 P^3-12 P^2 Q+11 P Q^2+4 P-3 Q^3-4 Q\right)} c_{-2}c_{1} + \\
&+\frac{8 P Q (2 P-3 Q)}{(2 P-Q) (2 P+Q) \left(4 P^3-12 P^2 Q+11 P Q^2+4 P-3 Q^3-4 Q\right)} c_{-1}c_{-1}c_{1} + \\
&+\frac{16 P Q}{(2 P+Q) \left(4 P^3-12 P^2 Q+11 P Q^2+4 P-3 Q^3-4 Q\right)} \psi_{-3/2}\psi_{-1/2}c_{1}+\dots\;,
\end{aligned}
\ee
\begin{equation}
\begin{aligned}\label{ReflectedOper2}
&\psi_{-3/2}^\mathrm{R}=-\frac{8 P^3+4 P^2 Q-10 P Q^2+8 P+3 Q^3+4 Q}{(2 P-Q) \left(4 P^2-8 P Q+3 Q^2+4\right)}\psi_{-3/2}+\\
&+\frac{16 i P Q}{(2 P-Q) \left(4 P^2-8 P Q+3 Q^2+4\right)}c_{-1}\psi_{-1/2}+\\
& +\frac{4 i \left(2 P^2 Q-P Q^2\right)}{(2 P+Q) \left(4 P^3-12 P^2 Q+11 P Q^2+4 P-3 Q^3-4 Q\right)} c_{-2} + \\
&+\frac{8 P Q}{(2 P+Q) \left(4 P^3-12 P^2 Q+11 P Q^2+4 P-3 Q^3-4 Q\right)} c_{-1}c_{-1}\psi_{1/2} -\\
&-\frac{16 P Q^2}{(P-Q) (2 P-Q) (2 P+Q) \left(4 P^2-8 P Q+3 Q^2+4\right)} \psi_{-3/2}\psi_{-1/2}\psi_{1/2}+\dots\;,\\
&c_{-2}^\mathrm{R}= -\frac{8 P^4-12 P^3 Q-6 P^2 Q^2+8 P^2+11 P Q^3-4 P Q-3 Q^4-4 Q^2}{(2 P-Q) \left(4 P^3-12 P^2 Q+11 P Q^2+4 P-3 Q^3-4 Q\right)} c_{-2} +\\
& +\frac{4 i P Q (2 P-3 Q)}{(2 P-Q) \left(4 P^3-12 P^2 Q+11 P Q^2+4 P-3 Q^3-4 Q\right)} c_{-1}c_{-1} +\\
&+ \frac{8 i P Q}{4 P^3-12 P^2 Q+11 P Q^2+4 P-3 Q^3-4 Q} \psi_{-3/2} \psi_{-1/2}+\dots\;,
\end{aligned}
\end{equation}
where dots stand for higher terms which do not contribute to the scalar products \eqref{ChainSpecBasis1}, \eqref{ChainSpecBasis2} on the considered below levels.

\paragraph{Basis elements in terms of the Uglov polynomials.}
The generators of $\cA(2,2)$ can be expressed in the following way \cite{Belavin:2012eg}
\begin{equation}\begin{aligned}\label{generatorsAndPhi}
&w_{n}=a_{2 n}^{(1)}+a_{2 n}^{(2)}\;, \quad c_{n}=\frac{a_{2 n}^{(1)}-a_{2 n}^{(2)}}{2}\;,\\
&\sum_{n}h_{n} z^{-2 n}=1+\frac{(-1)^{\sigma+1}}{4}\left(\exp \left(2 \phi^{(1)}\right)+\exp \left(2 \phi^{(2)}\right)+\exp \left(-2 \phi^{(1)}\right)+\exp \left(-2 \phi^{(2)}\right)\right)\;,\\
&\sum_{r} \chi_{r} z^{-2 r}=(-1)^{\sigma+1} \frac{i}{2 \sqrt{2}}\left(\exp \left(\phi^{(1)}+\phi^{(2)}\right)-\exp \left(-\phi^{(1)}-\phi^{(2)}\right)\right)\;,\\
&\sum_{r} \psi_{r} z^{-2 r}=\frac{i}{2 \sqrt{2}}\left(\exp \left(\phi^{(1)}-\phi^{(2)}\right)-\exp \left(-\phi^{(1)}+\phi^{(2)}\right)\right)\;,
\end{aligned}\end{equation}
where all exponents are normal ordered, $a_n^{(1,2)}$ are the generators of the two representations of the Heisenberg algebra
\be\label{CRa}
\left[a_n^{(k)},a_m^{(l)}\right]=n\delta_{l,k}\delta_{n+m,0}
\ee
and $\phi^{(1,2)}(z)$ are
\be\label{defPhi}
\phi^{(k)}(z)=\sum_{n} \frac{a_{2 n+1}^{(k)}}{-2 n-1} z^{-2 n-1}\;, \quad k=1,2\;.
\ee
The explicit form of the operators $a_n^{(1,2)}$ in terms of the bosonized $\cA(2,2)$ generators on the lower levels are given in Appendix~\ref{SpecialBas}.

In addition we need the reflected $a_n^{(1,2)}$ operators which we denote by $a_n^{(1,2)\mathrm{R}}$. These operators are related to $c_n^{\mathrm{R}}$ and $\psi_r^{\mathrm{R}}$ in the same way as $a_n^{(1,2)}$ to $c_n$ and $\psi_r$ in \eqref{generatorsAndPhi}.

The connection of the basis with the Uglov polynomials can be expressed \cite{Belavin:2011tb} as
\be\label{BasisConj}
\ba{l}
|\lambda,\varnothing\rangle=J^{(2)}_{\lambda}(x)|P\rangle\;,\\
|\varnothing,\lambda\rangle=J_\lambda^{(2)}(y)|P\rangle\;,
\ea
\ee	
where $J^{(2)}_{\lambda}(x)$ is the Uglov polynomial associated with the Young diagram $\lambda$, whose definition is given in Appendix~\ref{UglovPol}. The Uglov polynomials can be written in terms of the power-sum symmetric polynomials $p_k(x)=\sum_j x^k_j$, related to the operators $a_k^{(2)}$ and $a_k^{(2)\mathrm{R}}$ as
\be\label{polyOper}
\ba{l}
p_{2k+1}(x)=a^{(2)}_{-2k-1}\;,\quad
p_{2k}(x)=ib^{-1}a^{(2)}_{-2k}\;,\\
p_{2k+1}(y)=a^{(2)\mathrm{R}}_{-2k-1}\;,\quad
p_{2k}(y)=ib^{-1}a^{(2)\mathrm{R}}_{-2k}\;,
\ea
\ee
so that the Uglov polynomials can be expressed in terms of $w_n,h_n,\chi_r,c_n,\psi_r$ and $c_n^\mathrm{R}, \psi_r^\mathrm{R}$ using \eqref{generatorsAndPhi}.

Now using eqs.~\eqref{BasisConj}, \eqref{polyOper}, \eqref{generatorsAndPhi} and the relations \eqref{ReflectedOper1}, \eqref{ReflectedOper2} for the reflected operators one can construct the subclass of the basis in terms of the $\cA(2,2)$ generators on the lower levels.

\section{The basis for $\hat{c} = 1$}
\label{c1}
In this section we are going to show that the expansion of the chain vectors, eq.~\eqref{ChainVector}, in terms of the orthogonal basis leads to the AGT representation of the conformal block. Therefore we have to verify \eqref{ChainSpecBasis1} for all the elements of the basis. In the previous section we established the connection of the Uglov polynomials with the basis elements if one of the diagrams is empty. As in the bosonic case \cite{Belavin:2011js} we expect this connection holds for the whole basis if $\hat{c}=1$.

For the particular value of the central charge $\hat{c}=1$ (correspondingly $Q=0$) the equations \eqref{ReflectedOper1},~\eqref{ReflectedOper2} reduce to
\be
\ba{l}
\psi_{-1/2}^\mathrm{R}=-\psi_{-1/2}\;,\quad \psi_{-3/2}^\mathrm{R}=-\psi_{-3/2}\;,\\
c_{-1}^\mathrm{R}=-c_{-1}\;,\quad
c_{-2}^\mathrm{R}=-c_{-2}\;.
\ea
\ee
For the considered below levels, from 1/2 to 2, it implies 
\be
a_k^{(2)\mathrm{R}}=a_k^{(1)}\;.
\ee
Hence for $\hat{c}=1$ the AGT basis depends on two commuting generators $a_n^{(1)}$ and $a_n^{(2)}$. The natural assumption is that in this case, $\hat{c}=1$, the basis can be represented as a product of two Uglov polynomials such that
\be\label{C1orthBasisUglov}
|\vec{\lambda}\rangle=	J^{(2)}_{\lambda_1}(a_k^{(2)})J^{(2)}_{\lambda_2}(a_k^{(1)})|P\rangle\;.
\ee

Below we calculate the scalar products of the basis elements \eqref{C1orthBasisUglov} with the chain vectors \eqref{ChainVector} and test these results against the relation \eqref{ChainSpecBasis1} which is obtained directly from the AGT correspondence (see Appendix~\ref{ChainVectorsAndSpecBasis}). For the subclass of the basis with one empty diagram we use the form of the elements given by the eq.~\eqref{BasisConj} and calculate the products for the generic value of $\hat{c}$.
For our purposes we need to define the conjugation relations for the bosonized $\cA(2,2)$ generators. The standard form of the NS conjugation $L_n^{+}=L_{-n}$ and $G_r^{+}=G_{-r}$ fixes the conjugation rule for the generators $c_k,\psi_r$ and $\hat{P}$ :
\be\label{conj}
c_{k}^{+}=-c_{-n}\;, \quad \psi_r^{+}=-\psi_{-r}\;, \quad \hat{P}^{+}=-\hat{P}
\ee
and the conjugation relations for the rest generators are the following
\be
w_n^+=w_{-n}\;,\quad h_n^+=h_{-n}\;,\quad\chi_{r}^+=\chi_{-r}\;.
\ee

\paragraph{Level 1/2:} At this level the chain vector~\eqref{ChainVector} is
\be
|1/2\rangle = |1/2 \rangle^{SV}=\frac{1}{2\Delta}G_{-1/2}|P\rangle\;.
\ee
The coefficient in front of $G_{-1/2}$ is fixed by the normalization of the vector  $|1/2 \rangle^{SV}$:
\be
^{SV}\langle 1/2| 1/2 \rangle^{SV}=\frac{1}{2\Delta}\;.
\ee
This is easily seen to be consistent with~\eqref{ChainNorms}, namely
\be
\langle 1/2| 1/2 \rangle=\frac{1}{2\Delta}=2\left(Z_{\mathrm{vec}}^{\mathrm{sym}}(\vec{P}, ((1),\varnothing))+Z_{\mathrm{vec}}^{\mathrm{sym}}(\vec{P}, (\varnothing,(1)))\right)\;.
\ee
The scalar products of the chain vector with the basis vectors at this level are
\be
\langle(1),\varnothing|1/2\rangle=-\frac{\sqrt{2}}{2 P+Q}\;,
\ee\be
\langle\varnothing,(1)|1/2\rangle=\frac{\sqrt{2}}{2 P-Q}\;,
\ee
which is in agreement with the relation~\eqref{ChainSpecBasis1}.

\paragraph{Level 1:} The chain vector is
\be
|1\rangle_{3,4} = |1\rangle^{SV}_{3,4}+|1\rangle^{H\oplus\widehat{\text{sl}}(2)_2}_{3,4}=\beta_1^{SV}L_{-1}|P\rangle+\beta_1^{H\oplus\widehat{\text{sl}}(2)_2}w_{-1}|P\rangle\;,
\ee
where the coefficients are found using \eqref{chain} and \eqref{HSl22ChainCommRel}:
\be
\beta_1^{SV}=\frac{\Delta+\Delta_3-\Delta_4}{2\Delta}\;,\quad \quad \beta_1^{H\oplus\widehat{\text{sl}}(2)_2}=\frac{i(Q-\alpha_3)}{2}\;.
\ee
Hence, the complete expression for the chain vector at this level is
\be
|1\rangle_{3,4}=\frac{\Delta+\Delta_3-\Delta_4}{2\Delta}L_{-1}|P\rangle + \frac{i(Q-\alpha_3)}{2} w_{-1}|P\rangle
\ee
and the norm:
\be
\ba{l}
_{1,2}\langle 1 |1\rangle_{3,4}=\frac{(\Delta+\Delta_1-\Delta_2)(\Delta+\Delta_3-\Delta_4)}{2\Delta}-\alpha_1(Q-\alpha_3)=\\
=\sum_{\vec{\lambda}, N_{+}(\vec{\lambda})=1 \atop N_{-}(\vec{\lambda})=1} Z_{\mathrm{f}}^{\mathrm{sym}}\left(\mu_{i}, \vec{P}, \vec{\lambda}\right) Z_{\mathrm{vec}}^{\mathrm{sym}}(\vec{P}, \vec{\lambda})\;.
\ea
\ee
The scalar products of the special basis vectors with the chain vector $|1\rangle$ are 
\be
\ba{l}
\langle (2),\varnothing|1\rangle=\frac{(Q+2P-2P_3)^2-4P_4^2}{4 (Q+2P)}\;,\\
\langle (1,1),\varnothing|1\rangle=\frac{(Q+2P-2P_3)^2-4P_4^2}{4 (Q+2P)}\;,\\
\langle \varnothing,(2)|1\rangle=\frac{(Q-2P-P_3)^2-4 P_4^2}{4 (Q-2P)}\;,\\
\langle \varnothing, (1,1)|1\rangle=\frac{(Q-2P-P_3)^2-4 P_4^2}{4 (Q-2P)}\;,
\ea
\ee
which are consistent with the relation~\eqref{ChainSpecBasis1}.

\paragraph{Level 3/2:} The chain vector is
\be
\ba{l}
|3/2\rangle_{3,4}=|3/2\rangle^{SV}_{3,4}+|1/2\rangle^{SV}_{3,4}~|1\rangle^{H\oplus\widehat{\text{sl}}(2)_2}_{3,4}=\\
=\left(\beta_{3/2}^{SV}G_{-3/2}+\beta_{1,~1/2}^{SV}L_{-1}G_{-1/2}\right)|P\rangle + \frac{1}{2\Delta}\beta_1^{H\oplus\widehat{\text{sl}}(2)_2}G_{-1/2}w_{-1}|P\rangle\;.
\ea
\ee
Using the normalization $^{SV}\langle 0|0\rangle^{SV}=1$ and the recursion relation~\eqref{chain} one  gets the coefficients in the explicit form
\be
\ba{l}
\beta_{3/2}^{SV}=-\dfrac{2 \left(\Delta _3-\Delta _4\right)}{2 (\hat{c}-3) \Delta +\hat{c}+4 \Delta ^2}\;,\\
\beta_{1,~1/2}^{SV}=\dfrac{2 (\hat{c}-3) \Delta +2 \left(\Delta _3-\Delta _4\right) (\hat{c}+2 \Delta )+\hat{c}+4 \Delta ^2}{4 \Delta  \left(2 (\hat{c}-3) \Delta +\hat{c}+4 \Delta^2\right)}\;.
\ea
\ee
Again, the norm of the chain vectors $_{1,2}\langle 3/2| 3/2 \rangle_{3,4}$ can be expressed as in~\eqref{ChainNorms}:
\be
\ba{l}
_{1,2}\langle 3/2| 3/2 \rangle_{3,4}=\\
=\dfrac{\frac{2 \left(\Delta _1-\Delta _2\right) \left(2 (\hat{c}-3) \Delta +2 \left(\Delta _3-\Delta _4\right) (\hat{c}+2 \Delta )+\hat{c}+4 \Delta ^2\right)}{2 (\hat{c}-3) \Delta +\hat{c}+4 \Delta ^2}+2 \Delta +2 \Delta _3-2 \Delta _4+4 \alpha _1 \left(\alpha _3-Q\right)+1}{8 \Delta }=\\
=2\sum_{\vec{\lambda}, N_{+}(\vec{\lambda})=1 \atop N_{-}(\bar{\lambda})=2} Z_{\mathrm{f}}^{\mathrm{sym}}\left(\mu_{i}, \vec{P}, \vec{\lambda}\right) Z_{\mathrm{vec}}^{\mathrm{sym}}(\vec{P}, \vec{\lambda})\;.
\ea
\ee
The special basis at this level consists of four vectors, labeled by one empty and one non-empty diagram, and four vectors with both non-empty diagrams. Taking the scalar products of the first four vectors with the chain vector $|3/2\rangle$ one gets for generic value of the central charge  $c(b)$:
\be
\ba{l}
\langle (1,1,1),\varnothing | 3/2\rangle = -\frac{\sqrt{2} \left(b+\frac{1}{2} \left(b+\frac{1}{b}\right)+P-P_3-P_4\right) \left(b+\frac{1}{2} \left(b+\frac{1}{b}\right)+P-P_3+P_4\right)}{\left(b+\frac{1}{b}+2 P\right) \left(3 b+\frac{1}{b}+2 P\right)}\;,\\
\langle (3),\varnothing | 3/2\rangle = -\frac{\sqrt{2} \left(\frac{1}{2} \left(b+\frac{1}{b}\right)+\frac{1}{b}+P-P_3-P_4\right) \left(\frac{1}{2} \left(b+\frac{1}{b}\right)+\frac{1}{b}+P-P_3+P_4\right)}{\left(b+\frac{1}{b}+2 P\right) \left(b+\frac{3}{b}+2 P\right)}\;,\\
\langle \varnothing, (1,1,1) | 3/2\rangle = -\frac{\sqrt{2} \left(b+\frac{1}{2} \left(b+\frac{1}{b}\right)-P-P_3-P_4\right) \left(b+\frac{1}{2} \left(b+\frac{1}{b}\right)-P-P_3+P_4\right)}{\left(b+\frac{1}{b}-2 P\right) \left(3 b+\frac{1}{b}-2 P\right)}\;,\\
\langle \varnothing, (3) | 3/2\rangle = -\frac{\sqrt{2} \left(\frac{1}{2} \left(b+\frac{1}{b}\right)+\frac{1}{b}-P-P_3-P_4\right) \left(\frac{1}{2} \left(b+\frac{1}{b}\right)+\frac{1}{b}-P-P_3+P_4\right)}{\left(b+\frac{1}{b}-2 P\right) \left(b+\frac{3}{b}-2 P\right)}\;,
\ea
\ee
which gives \eqref{ChainSpecBasis1}.\newline
For the remaining four vectors the scalar products with the chain vector can be obtained explicitly only for $\hat{c} = 1$, or $b=i$: 
\be
\ba{l}
\langle (2),(1)| 3/2 \rangle = \frac{(P-P_3-P_4-i) (P-P_3+P_4-i)}{\sqrt{2} (2 P-2 i) P}\;,\\
\langle (1,1),(1)| 3/2 \rangle = \frac{(P-P_3-P_4+i) (P-P_3+P_4+i)}{\sqrt{2} P (2 P+2 i)}\;,\\
\langle (1),(2)| 3/2 \rangle = \frac{(-P-P_3-P_4-i) (-P-P_3+P_4-i)}{\sqrt{2} P (2 P+2 i)}\;,\\
\langle (1),(1,1)| 3/2 \rangle = \frac{(-P-P_3-P_4+i) (-P-P_3+P_4+i)}{\sqrt{2} (2 P-2 i) P}\;,
\ea
\ee
which are also consistent with~\eqref{ChainSpecBasis1}.

The details of the similar calculations for the level two are collected in Appendix~\ref{Level2}. These checks confirm the relation~\eqref{ChainSpecBasis1} between the chain vectors and the elements of the basis proposed in~\cite{Belavin:2012eg} and allow one to connect the correlation functions in  $\cN=1$  super-Virasoro CFT 
with the Uglov polynomials, for $\hat{c} = 1$.

\section{Concluding remarks}
\label{concl}

In this paper we studied the properties of the AGT basis in the case of  $\cN=1$ super-Virasoro CFTs. To this end we considered an explicit construction of the four-point correlation function on the sphere in terms  of the AGT basis elements. The standard CFT construction is based on the operator product expansion which can be formulated in terms of the chain vectors, encoding the contribution of the descendents in the OPEs of the primary fields. The AGT correspondence assumes the consideration of the extended  theory, obeying $A(2,2)$ chiral algebra, which contains the $\cN=1$ super-Virasoro (NSR algebra) as a subalgebra. The construction of the correlation functions can be then reduced to the analysis of the relations between the chain vectors and the AGT basis elements, $\{|P\rangle_{\lambda_1^\sigma,\lambda_2^\sigma}\}$, which arises in the $\cA(2,2)$ case.   

In~\cite{Belavin:2012eg} a connection between the AGT basis and the Uglov polynomials was proposed for the subclass of the basis elements of the form $|P\rangle_{\lambda_1^\sigma,\emptyset^\sigma}$; it was argued there that  the elements $|P\rangle_{\emptyset^\sigma, \lambda_2^\sigma}$ can be obtained using the second possible bosonization of the NSR algebra. Application of these results to the construction of the correlation functions  involves a number of issues and requires an analysis performed in this paper.

In order to construct the correlation function we established a connection between the chain vectors and the elements of the special basis. This consideration relays on the AGT correspondence for the $\cN=1$ super-Virasoro theory~\cite{Belavin:2011tb} without reference to the explicit form of the special basis. Due to the orthogonality of the special basis one can find explicitly the scalar products between the chain vectors and the basis elements. The validity of these relations ensures the correct result for the correlation function and represents the main check of the correct form of the basis elements. Note that these relations themselves are not sufficient to derive the basis elements.      

The next step is to write the basis elements in terms of the bosonized generators. The procedure for obtaining the subclass $|P\rangle_{\lambda_1^\sigma,\emptyset^\sigma}$ is described in~\cite{Belavin:2012eg} and requires computation of the Uglov polynomials, for which we used the algorithm described in detail in Appendix~\ref{UglovPol}. This procedure fits our present consideration up to normalizations and conventions, which we changed for the sake of consistency with the found relations with the chain vectors.    
In order to construct the subclass  of the elements with the second non-empty diagram, $|P\rangle_{\emptyset^\sigma, \lambda_2^\sigma}$, it is necessary to establish the connection between the reflected generators of the bosonized NS algebra, $c_k^\mathrm{R}, \psi_r^\mathrm{R}$, and the non-reflected ones, $c_k, \psi_r$. This relation follows from the definition of the two bosonizations and the solution can be obtained as an expansion in terms of the modes.  This allows one to express the chain vectors in terms of the $\cA(2, 2)$ generators and to obtain their explicit form on each particular level of the descendants contribution.  We find that the problem of the appropriately fixed normalization and the proper choice of the conjugation rule for the basis elements play an important role. In particular, the correct choice of the normalization is crucial for the analysis when both diagrams are not empty.

In this paper we also considered the special value of the central charge $\hat{c} = 1$. We showed that in this case  the elements of the special basis can be written in terms of products of two Uglov polynomials with the variables expressed in terms of the bosonized  generators by means of two different Feigin-Fuchs bosonizations of the NSR algebra. It provides an explicit check on the level of correlation functions of the conjecture about the form of the basis in the  case $\hat{c} = 1$, stated in~\cite{Belavin:2012eg}.

\vspace{7mm}
\noindent \textbf{Acknowledgements.} We are grateful to M.~Lashkevich for useful comments on the draft of this article. A.~Zh.  thanks the DESY Hamburg for  hospitality  during the program ``YRISW 2020: A modern primer for superconformal field theories''.

\appendix
\section{Uglov polynomials}
\label{UglovPol}
\paragraph{Notation.}
Given a partition $\lambda=(\lambda_1,\lambda_2,\dots)$ we are going to call $l(\lambda)$ its length i.e. the number of non-zero elements $\lambda_1,\lambda_2,\dots$ in $\lambda$. Also we are going to identify a partition $\lambda$ with its diagram, which can be defined as a set of squares with coordinates $(i,j)$:
\be
\left\{(i, j) | 1 \leq i \leq l(\lambda), 1 \leq j \leq \lambda_{i}\right\}\;,
\ee
where $i$ is increasing downward and $j$ is increasing from left to right. For example, below is written the diagram for the partition $\lambda=(5,3,2,1)$
\be
\tableau{5 3 2 1}\;.
\ee
We denote by $a_\lambda(s)$ and $l_\lambda(s)$  correspondingly the arm length of the square $s\in\lambda$ and its leg length. The arm length $a_\lambda(s)$ and the leg length $l_\lambda(s)$ are the numbers of squares of the diagram $\lambda$ to the east and south from the square $s$ respectively. For example, in the diagram above the square $(1,1)$ has an arm length $a_\lambda(s)=4$ and a leg length $l_\lambda(s)=3$.

Having two partitions $\mu$ and $\lambda$ of the equal number of squares $|\mu|=|\lambda|$ we will write $\lambda>\mu$ if there exist an integer $k$ such that $\lambda_k>\mu_k$. For example, $\lambda=(4,2)>\mu=(4,1,1)$:
\be
\tableau{4 2} > \tableau{4 1 1}\;.
\ee

Given two partitions $\mu$ and $\lambda$ such that $\mu \subset \lambda$ the set-theoretic difference $\theta=\lambda-\mu$ is called a skew diagram. The skew diagram $\theta$ is called a horizontal strip if it has at most one square in each column. For example, for diagrams $\lambda=(4,4,1)$ and $\mu=(4,1)$ the skew diagram 
\be
\tableau{4 4 1} - \tableau{4 1}
\ee
is a horizontal strip.

For a partition $\lambda$ a tableau $T$ of shape $\lambda$ is a sequence of partitions:
\be
\emptyset=\lambda^{(0)} \subset \lambda^{(1)} \subset \cdots \subset \lambda^{(r)}=\lambda
\ee
such that each skew diagram $\theta^{(i)}=\lambda^{(i)}-\lambda^{(i-1)}(1 \leq i \leq r)$ is a horizontal strip. The sequence $\left(\left|\theta^{(1)}\right|, \ldots,\left|\theta^{(r)}\right|\right)$ is called the weight of $T$. There can be a few tableaux of the same shape and with the same weight. For example, the tableaux $T_1$ and $T_2$ of shape $\lambda=(3,1)$ with the weight $(2,1,1)$ are the following sequences of diagrams:
\be
T_1=\left(\varnothing,\tableau{1},\tableau{2},\tableau{3 1}\right)\;,\quad
T_2=\left(\varnothing,\tableau{1},\tableau{1 1},\tableau{3 1}\right)\;.
\ee

\paragraph{Macdonald polynomials.}
We consider  $\mathbf{C}(q,t)$-algebra of symmetric polynomials in variables $x_1,\dots,x_n$ $\Lambda_{n}^{q, t}=\left(\mathbf{C}(q, t)\left[x_{1}, \ldots, x_{n}\right]\right)^{S_{n}}$. The scalar product in $\Lambda_{n}^{q, t}$ is defined as follows
\be\label{ScalarProducMacdonald}
\langle f, g\rangle_{q, t}=\frac{1}{n !} \prod_{j=1}^{n} \int \frac{d w_{j}}{2 \pi i w_{j}} \prod_{1 \leq i \neq j \leq n} \frac{\left(w_{i} w_{j}^{-1} ; q\right)_{\infty}}{\left(t w_{i} w_{j}^{-1} ; q\right)_{\infty}} \overline{f\left(w_{1}, \ldots, w_{n}\right)} g\left(w_{1}, \ldots, w_{n}\right)
\ee
for $f=f(x_1,\dots,x_n), g=g(x_1,\dots,x_n)\in\Lambda_{n}^{q, t}$. In the equation \eqref{ScalarProducMacdonald} the integration in each of the variables $w_j$ is taken along the unit circle in the complex plane, and $(x ; q)_{\infty}=\prod_{r=0}^{\infty}\left(1-x q^{r}\right)$. For each partition $\lambda$ of length less or equal to $n$ the Macdonald Polynomial $P_{\lambda}(q, t)=P_{\lambda}\left(x_{1}, \dots, x_{n} ; q, t\right)$ is uniquely defined element of $\Lambda_{n}^{q, t}$ such that 
\be
P_{\lambda}(q, t)=m_{\lambda}+\sum_{\mu<\lambda} u_{\lambda \mu}(q, t) m_{\mu}\;,
\ee
where $u_{\lambda \mu}(q, t) \in \mathbf{C}(q, t)$ and $m_{\lambda}=m_{\lambda}\left(x_{1}, \dots, x_{n}\right)$ is the monomial symmetric polynomial, defined as
\be
m_\lambda(x_1,\dots,x_n)=\text{Sym}\left(x_1^{\lambda_1}\dots x_n^{\lambda_n}\right)\;.
\ee
In the last equation $Sym$ means symmetrization, for example $m_{(2,1)}(x_1,x_2,x_3)=x_1^2 x_2 + + x_1^2 x_3 + x_2^2 x_1 + x_2^2 x_3 + x_3^2 x_1 +  x_3^2 x_2$.
The coefficient of $m_\mu$ in the expansion of $P_\lambda(q, t)$ is
\be
u_{\lambda \mu}(q, t)=\sum_{T} \psi_{T}(q, t)
\ee
summed over the tableaux of shape $\lambda$ and weight $\mu$. 
For given tableau  $T=(\lambda^{(0)},\dots,\lambda^{(r)})$ one can define a function $\psi_T(q,t)$
\be
\psi_{T}(q, t)=\prod_{i=1}^{r} \psi_{\lambda^{(i)} / \lambda^{(i-1)}}(q, t)\;.
\ee
To define the function $\psi_{\lambda / \mu}$ for two partitions  $\mu$ and $\lambda$ such that $\lambda / \mu$ is a horizontal strip, let $R_{\lambda / \mu}$ denote the union of the rows of $\mu$ that intersects $\lambda / \mu$. Then
\be
\psi_{\lambda / \mu}(q, t)=\prod_{s \in R_{\lambda / \mu}} \frac{b_{\mu}(s ; q, t)}{b_{\lambda}(s ; q, t)}\;,
\ee
where 
\be
b_{\lambda}(s ; q, t)=\left\{\begin{array}{ll}
	\dfrac{1-q^{a_{\lambda}(s)} t^{l_{\lambda}(s)+1}}{1-q^{a_{\lambda}(s)+1} t^{l_{\lambda}(s)}}\;, &\quad \text {if } s \in \lambda \\
	1\;, &\quad \text {otherwise }
\end{array}\right.
\ee
For each partition $\lambda$ one can define another set of Macdonald polynomials
\be
J_{\lambda}(q, t)=c_{\lambda}(q, t) P_{\lambda}(q, t)\;,
\ee
where 
\be
c_{\lambda}(q, t)=\prod_{s \in \lambda}\left(1-q^{a(s)} t^{l(s)+1}\right)
\ee
is an additional factor such that the coefficients $v_{\lambda \mu}$ in the expansion
\be
J_{\lambda}(q, t)=\sum_{\mu \leq \lambda} v_{\lambda, \mu}(q, t) m_{\mu}
\ee
are polynomials on $q$ and $t$ with integer coefficients. This basis is orthogonal with the norm
\be
\left\langle J_{\lambda}(q, t), J_{\lambda}(q, t)\right\rangle_{q, t}=\prod_{s \in \lambda}\left(1-q^{a(s)} t^{l(s)+1}\right)\left(1-q^{a(s)+1} t^{l(s)}\right)\;.
\ee

\paragraph{Uglov polynomials.}
Let now $\alpha$ be a positive real number and $\omega_{p}=\exp \left(\frac{2 \pi i}{p}\right)$ and consider the limit:
\be
q=\omega_{p} \rho^\alpha\;, \quad t=\omega_{p} \rho\;, \quad \rho \rightarrow 1\;.
\ee 
It was shown in \cite{Uglov:1997ia} that this limit is well defined for Macdonald polynomials $P_{\lambda}(q, t)$. This limit is denoted as rang $p$ Uglov polynomials $P^{(\alpha,p)}_\lambda$. One can show that the basis of the Uglov polynomials with the integer normalization $J_{\lambda}^{(\alpha, p)}$ can be defined by (for more details see \cite{Belavin:2012eg})
\be
J_{\lambda}^{(\alpha, p)}=\lim _{\tau \rightarrow 0}\left(\frac{J_{\lambda}(q, t)}{\tau^{|\lambda^{\diamond}|}\epsilon_1^{|\lambda^{\diamond}|} \prod_{s \in \lambda-\lambda^\diamond}\left(1-\omega_{p}^{a_{\lambda}(s)+l_{\lambda}(s)+1}\right)}\right)=P_{\lambda}^{(\alpha, p)} \prod_{s \in \lambda^\diamond}\left(l_{\lambda}(s)+1-\alpha a_{\lambda}(s)\right)\;,
\ee
where $\lambda^{\diamond}=\left\{s \in \lambda | a_{\lambda}(s)+l_{\lambda}(s)+1 \equiv 0 \bmod 2\right\}$. 
In this work we work only with parameters $\alpha=-1/b^2$ and $p=2$.
We follow \cite{Belavin:2012eg} and use $J_{\lambda}^{(2)}$ defined as
\be
J_{\lambda}^{(2)}=\lim _{\tau \rightarrow 0}\left(\frac{(-1)^{n(\lambda)}}{\tau^{|\lambda^\diamond|} 2^{|\lambda|-|\lambda^\diamond|}} J_{\lambda}(q, t)\right)\;.
\ee
We use~\cite{sagemath} to get Macdonald polynomials $J_{\lambda}(q, t)$ and then find the limit according to the equation above. Denoting $p_\lambda=\prod_i p_{\lambda_{i}}$ we get:
\be
\ba{l}\label{Uglovs}
J_{(1)}^{(2)}=p_{1}\;,\\
J_{(1,1)}^{(2)}=b p_{1,1}-b p_2\;,\quad J_{(2)}^{(2)}=\dfrac{p_{1,1}}{b}-b p_2\;,\\
J_{(1,1,1)}^{(2)}=-b p_{2,1}+\dfrac{1}{3} b p_{1,1,1}+\dfrac{2}{3} b p_3\;,\\
J_{(2,1)}^{(2)}=\dfrac{p_3}{3}-\dfrac{1}{3} p_{1,1,1}\;,\\
J_{(3)}^{(2)}=-b p_{2,1}+\dfrac{p_{1,1,1}}{3 b}+\dfrac{2 p_3}{3 b}\;,\\
J_{(4)}^{(2)}=b^2 p_{2,2}+\dfrac{8 p_{3,1}}{3 b^2}+\dfrac{p_{1,1,1,1}}{3 b^2}-2 p_{2,1,1}-2 p_4\;,\\
J_{(3,1)}^{(2)}=b^2 p_{2,2}-b^2 p_{2,1,1}+\dfrac{2 p_{3,1}}{3 b^2}-\dfrac{2 p_{1,1,1,1}}{3 b^2}+\dfrac{2}{3} p_{3,1}+p_{2,1,1}+\dfrac{1}{3} p_{1,1,1,1}-2 p_{4}\;,\\
J_{(2,2)}^{(2)}=b^2 p_{2,2}+\dfrac{4}{3} p_{3,1}-\dfrac{1}{3} p_{1,1,1,1}-(b^2+1) p_{4}\;,\\
J_{(2,1,1)}^{(2)}=b^2 p_{2,2}+\dfrac{2}{3}( b^2+1) p_{3,1}+(b^2-1) p_{2,1,1}-\dfrac{2}{3} b^2 p_{1,1,1,1}+\dfrac{1}{3} p_{1,1,1,1}-2 b^2 p_4\;,\\
J_{(1,1,1,1)}^{(2)}=b^2 p_{2,2}+\dfrac{8}{3} b^2 p_{3,1}-2 b^2 p_{2,1,1}+\dfrac{1}{3} b^2 p_{1,1,1,1}-2 b^2 p_{4}\;.
\ea
\ee

\section{The explicit form of the special basis}\label{SpecialBas}
In this appendix we explicitly write the elements of the orthogonal basis at the levels from $1/2$ to $2$ and check that their scalar products obey the AGT relation~\eqref{AGTrel}.
We can summarize equations~\eqref{BasisConj},~\eqref{polyOper} and~\eqref{generatorsAndPhi} that we are using in order to express the Uglov polynomials in terms of $\mathcal{A}(2,2)$ generators in the following table. Here we use the connection between the parameter $\sigma$ arising in~ \eqref{generatorsAndPhi} and the number of indices $k$ of the operators in the LHS: $\sigma=k~ \mod 2$.
\begin{center}
\begin{tabular}{|c|c|}
	\hline
	\textbf{Level 1/2} &  \\
	\hline
	$a_{-1}^{(2)}|P\rangle$ & $-\frac{i}{\sqrt{2}}\left(\chi_{-1/2}-\psi_{-1/2}\right)|P\rangle$ \\
	\hline
	\textbf{Level 1} &  \\
	\hline
	$a_{-2}^{(2)}|P\rangle$ & $(\frac{1}{2}w_{-1}-c_{-1})|P\rangle$ \\
	\hline
	$a_{-1}^{(2)}a_{-1}^{(2)}|P\rangle$ & $-\left(\frac{1}{2}h_{-1}+\chi_{-1/2}\psi_{-1/2}\right)|P\rangle$ \\
	\hline
	\textbf{Level 3/2} &  \\
	\hline
	$\left(a_{-1}^{(2)}a_{-1}^{(2)}a_{-1}^{(2)}+2a_{-3}^{(2)}\right)|P\rangle$ & $\frac{-3i}{\sqrt{2}}\left(\chi_{-3/2}-\psi_{-3/2}-\frac{1}{2}h_{-1}\left(\chi_{-1/2}-\psi_{-1/2}\right)\right)|P\rangle$  
	\\
	\hline
	$a_{-2}^{(2)}a_{-1}^{(2)}|P\rangle$ & $-\frac{i}{\sqrt{2}}\left(\chi_{-1/2}-\psi_{-1/2}\right)(\frac{1}{2}w_{-1}-c_{-1})|P\rangle$ \\
	\hline
	$a_{-2}^{(2)}a_{-1}^{(1)}|P\rangle$ & $-\frac{i}{\sqrt{2}}\left(\chi_{-1/2}+\psi_{-1/2}\right)(\frac{1}{2}w_{-1}-c_{-1})|P\rangle$ \\
	\hline
	$a_{-1}^{(2)}a_{-1}^{(2)}a_{-1}^{(1)}|P\rangle$ & $-\frac{i}{\sqrt{2}}\left(\chi_{-3/2}+\psi_{-3/2}+\frac{1}{2}h_{-1}\left(\chi_{-1/2}+\psi_{-1/2}\right)\right)|P\rangle$  \\
	\hline
	\textbf{Level 2} &  \\
	\hline
	$a_{-4}^{(2)}|P\rangle$ & $(\frac{1}{2}w_{-2}-c_{-2})|P\rangle$ \\
	\hline
	$a_{-3}^{(2)}a_{-1}^{(2)}|P\rangle$ & $\frac{1}{2}\left(\frac{1}{4}h_{-1}^2+h_{-1}\psi_{-1/2}\chi_{-1/2}+\chi_{-3/2}\chi_{-1/2}+\psi_{-3/2}\psi_{-1/2}-\right.$\\
	& $\left.-\left(\frac{1}{2}h_{-2}+\chi_{-3/2}\psi_{-1/2}+\psi_{-3/2}\chi_{-1/2}\right)\right)|P\rangle$  \\
	\hline
	$a_{-2}^{(2)}a_{-2}^{(2)}|P\rangle$ & $(\frac{1}{2}w_{-1}-c_{-1})(\frac{1}{2}w_{-1}-c_{-1})|P\rangle$ \\
	\hline
	$a_{-2}^{(2)}a_{-1}^{(2)}a_{-1}^{(2)}|P\rangle$ & $-\left(\frac{1}{2}h_{-1}+\chi_{-1/2}\psi_{-1/2}\right)(\frac{1}{2}w_{-1}-c_{-1})|P\rangle$ \\
	\hline
	$a_{-1}^{(2)}a_{-1}^{(2)}a_{-1}^{(2)}a_{-1}^{(2)}|P\rangle$ & $-\left(\frac{1}{4}h_{-1}^2+h_{-1}\psi_{-1/2}\chi_{-1/2}+\chi_{-3/2}\chi_{-1/2}+\psi_{-3/2}\psi_{-1/2}+\right.$\\
	& $\left.+2\left(\frac{1}{2}h_{-2}+\chi_{-3/2}\psi_{-1/2}+\psi_{-3/2}\chi_{-1/2}\right)\right)|P\rangle$  \\
	\hline
	$\left(a_{-3}^{(2)}a_{-1}^{(1)}-(a_{-1}^{(2)})^3a_{-1}^{(1)}\right)|P\rangle$ & $\frac{3}{2}\left(\chi_{-3/2}+\psi_{-3/2}\right)\left(\chi_{-1/2}-\psi_{-1/2}\right)|P\rangle$  \\
	\hline
	$a_{-2}^{(2)}a_{-2}^{(1)}|P\rangle$ & $(\frac{1}{2}w_{-1}-c_{-1})(\frac{1}{2}w_{-1}+c_{-1})|P\rangle$ \\
	\hline
	$a_{-1}^{(2)}a_{-1}^{(2)}a_{-2}^{(1)}|P\rangle$ & $-\left(\frac{1}{2}h_{-1}+\chi_{-1/2}\psi_{-1/2}\right)(\frac{1}{2}w_{-1}+c_{-1})|P\rangle$  \\
	\hline
	$a_{-1}^{(2)}a_{-1}^{(2)}a_{-1}^{(1)}a_{-1}^{(1)}|P\rangle$ & $\left(\frac{1}{4}h_{-1}^2-\chi_{-3/2}\chi_{-1/2}-\psi_{-3/2}\psi_{-1/2}\right)|P\rangle$ \\
	\hline
\end{tabular}
\end{center}
\paragraph{Level 1/2:}
\be
\ba{l}
|(1),\varnothing\rangle=\frac{-i}{2}\left(\chi_{-1/2}-\psi_{-1/2}\right)|P\rangle\;,\\
|\varnothing,(1)\rangle=\frac{-i}{2}\left(\chi_{-1/2}-\psi^{\mathrm{R}}_{-1/2}\right)|P\rangle\;.
\ea
\ee
The matrix of the scalar products at this level coincide with the AGT relation~\eqref{AGTrel}:
\be
\begin{pmatrix}
	-\frac{2 P}{Q+2P} & 0 \\
	0 & \frac{2 P}{Q-2P}
\end{pmatrix}
=
\begin{pmatrix}
	\frac{Z_{\mathrm{bif}}\left(0|P,((1)^1,(\varnothing)^1),P,((1)^1,(\varnothing)^1)\right)}{\Omega_{(1)}^2(P)} & \frac{Z_{\mathrm{bif}}\left(0|P,((1)^1,(\varnothing)^1),P,((\varnothing)^1,(1)^1)\right)}{\Omega_{(1)}(P)\Omega_{(1)}(-P)} \\
	\frac{Z_{\mathrm{bif}}\left(0|P,((\varnothing)^1,(1)^1),P,((1)^1,(\varnothing)^1)\right)}{\Omega_{(1)}(P)\Omega_{(1)}(-P)} & \frac{Z_{\mathrm{bif}}\left(0|P,((\varnothing)^1,(1)^1),P,((\varnothing)^1,(1)^1)\right)}{\Omega_{(1)}^2(-P)}
\end{pmatrix}.
\ee
\paragraph{Level 1:}
\be
\ba{l}
|(1,1),\varnothing\rangle=\left[-b \left(\frac{1}{2}h_{-1}+\chi_{-1/2}\psi_{-1/2}\right)-i \left(\frac{1}{2}w_{-1}-c_{-1}\right)\right]|P\rangle\;,\\
|(2),\varnothing\rangle=\left[-b^{-1} \left(\frac{1}{2}h_{-1}+\chi_{-1/2}\psi_{-1/2}\right)-i \left(\frac{1}{2}w_{-1}-c_{-1}\right)\right]|P\rangle\;,\\
|\varnothing,(1,1)\rangle=\left[-b \left(\frac{1}{2}h_{-1}+\chi_{-1/2}\psi^{\mathrm{R}}_{-1/2}\right)-i \left(\frac{1}{2}w_{-1}-c^{\mathrm{R}}_{-1}\right)\right]|P\rangle\;,\\
|\varnothing,(2)\rangle=\left[-b^{-1} \left(\frac{1}{2}h_{-1}+\chi_{-1/2}\psi^{\mathrm{R}}_{-1/2}\right)-i \left(\frac{1}{2}w_{-1}-c^{\mathrm{R}}_{-1}\right)\right]|P\rangle\;.
\ea
\ee
The matrix of the scalar products at this level agrees with the AGT relation~\eqref{AGTrel}:
\be
\ba{c}
\mathrm{diag}\left(-\frac{4 \left(b^2-1\right) P}{b^2 (2 P+Q)},\frac{4 \left(b^2-1\right) P}{2 P+Q},-\frac{4 \left(b^2-1\right) P}{b^2 (2 P-Q)},\frac{4 \left(b^2-1\right) P}{2 P-Q}\right)
=\frac{Z_{\mathrm{bif}}\left(0|P,\vec{\lambda}^0,P,\vec{\mu}^0\right)}{\Omega_{\lambda_1,\lambda_2}(P)\Omega_{\mu_1,\mu_2}(P)}\;.
\ea
\ee
\paragraph{Level 3/2:}
The special basis at this level consists of 4 vectors with one empty diagram and 4 vectors with both diagrams being non-empty. The vectors with one empty diagram are (for this vectors we perform the computation for generic value of $Q$):
\be
\begin{aligned}
	|(1,1,1),\varnothing\rangle=&\left[\frac{-ib}{\sqrt{2}}\left(\chi_{-3/2}-\psi_{-3/2}\right)-\frac{1}{\sqrt{2}}\left(\frac{1}{2}w_{-1}-c_{-1}\right)\left(\chi_{-1/2}-\psi_{-1/2}\right)+\right.\\
	&\left.+\frac{ib}{2\sqrt{2}}h_{-1}\left(\chi_{-1/2}-\psi_{-1/2}\right)\right]|P\rangle\;,\\
	|(3),\varnothing\rangle=&\left[\frac{-i}{\sqrt{2}b}\left(\chi_{-3/2}-\psi_{-3/2}\right)-\frac{1}{\sqrt{2}}\left(\frac{1}{2}w_{-1}-c_{-1}\right)\left(\chi_{-1/2}-\psi_{-1/2}\right)+\right.\\
	&\left.+\frac{i}{2\sqrt{2} b}h_{-1}\left(\chi_{-1/2}-\psi_{-1/2}\right)\right]|P\rangle\;,\\
	|\varnothing,(1,1,1)\rangle=&\left[\frac{-ib}{\sqrt{2}}\left(\chi_{-3/2}-\psi_{-3/2}^{\mathrm{R}}\right)-\frac{1}{\sqrt{2}}\left(\frac{1}{2}w_{-1}-c_{-1}^{\mathrm{R}}\right)\left(\chi_{-1/2}-\psi_{-1/2}^{\mathrm{R}}\right)+\right.\\
	&\left.+\frac{ib}{2\sqrt{2}}h_{-1}\left(\chi_{-1/2}-\psi_{-1/2}^{\mathrm{R}}\right)\right]|P\rangle\;,\\
	|\varnothing,(3)\rangle=&\left[\frac{-i}{\sqrt{2}b}\left(\chi_{-3/2}-\psi_{-3/2}^{\mathrm{R}}\right)-\frac{1}{\sqrt{2}}\left(\frac{1}{2}w_{-1}-c_{-1}^{\mathrm{R}}\right)\left(\chi_{-1/2}-\psi_{-1/2}^{\mathrm{R}}\right)+\right.\\
	&\left.+\frac{i}{2\sqrt{2} b}h_{-1}\left(\chi_{-1/2}-\psi_{-1/2}^{\mathrm{R}}\right)\right]|P\rangle\;.
\end{aligned}
\ee
The matrix of the scalar products for these vectors is the following
\be
\ba{c}
\mathrm{diag}\left(	\frac{8 \left(b^2-1\right) P (b P+1)}{b \left(b^2+2 b P+1\right) \left(b^2+2 b P+3\right)},-\frac{8 b^2 \left(b^2-1\right) P (b+P)}{\left(b^2+2 b P+1\right) \left(3 b^2+2 b P+1\right)},\frac{8 \left(b^2-1\right) P (b P-1)}{b \left(b^2-2 b P+1\right) \left(b^2-2 b P+3\right)},\frac{8 b^2 \left(b^2-1\right) P (b-P)}{\left(b^2-2 b P+1\right) \left(3 b^2-2 b P+1\right)}\right)
\ea
\ee
and it coincides with the AGT relation \eqref{AGTrel}:
\be
\frac{Z_{\mathrm{bif}}\left(0|P,\vec{\lambda}^0,P,\vec{\mu}^0\right)}{\Omega_{\lambda_1,\lambda_2}(P)\Omega_{\mu_1,\mu_2}(P)}\;.
\ee

The vectors of the special basis with both non-empty diagrams are the following (for these vectors we assume $Q=0$):
\be
\begin{aligned}
	|(2),(1)\rangle=J^{(2)}_{(2)}(a_k^{(2)})J^{(2)}_{(1)}(a_k^{(1)})|P\rangle=-i\left(a_{-1}^{(2)}a_{-1}^{(2)}+a_{-2}^{(2)}\right)a_{-1}^{(1)}|P\rangle\;,\\
	|(1,1),(1) \rangle =J^{(2)}_{(1,1)}(a_k^{(2)})J^{(2)}_{(1)}(a_k^{(1)})|P\rangle=i\left(a_{-1}^{(2)}a_{-1}^{(2)}-a_{-2}^{(2)}\right)a_{-1}^{(1)}|P\rangle \;,\\
	|(1),(2)\rangle =J^{(2)}_{(1)}(a_k^{(2)})J^{(2)}_{(2)}(a_k^{(1)})|P\rangle =-ia_{-1}^{(2)}\left(a_{-1}^{(1)}a_{-1}^{(1)}+a_{-2}^{(1)}\right)|P\rangle\;, \\
	|(1),(1,1)\rangle =J^{(2)}_{(1)}(a_k^{(2)})J^{(2)}_{(1,1)}(a_k^{(1)})|P\rangle=ia_{-1}^{(2)}\left(a_{-1}^{(1)}a_{-1}^{(1)}-a_{-2}^{(1)}\right)|P\rangle\;.
\end{aligned}
\ee
Using the definition \eqref{C1orthBasisUglov} and the expressions in the table above one can write these vectors in the following way:
\be
\begin{aligned}
	|(2),(1)\rangle=& -\frac{1}{\sqrt{2}}\left[\left(\frac{w_{-1}}{2}-c_{-1}\right) \left(\chi_{-\frac{1}{2}}+\psi_{-\frac{1}{2}}\right)+\right.\\
	&\left.+\left(\frac{1}{2} h_{-1} \left(\chi_{-\frac{1}{2}}+\psi_{-\frac{1}{2}}\right)+\chi_{-\frac{3}{2}}+\psi_{-\frac{3}{2}}\right)\right]|P\rangle\;,\\
	|(1,1),(1) \rangle =& \frac{1}{\sqrt{2}}\left[-\left(\frac{w_{-1}}{2}-c_{-1}\right) \left(\chi_{-\frac{1}{2}}+\psi_{-\frac{1}{2}}\right)+\right.\\
	&\left.+\left(\frac{1}{2} h_{-1} \left(\chi_{-\frac{1}{2}}+\psi_{-\frac{1}{2}}\right)+\chi_{-\frac{3}{2}}+\psi_{-\frac{3}{2}}\right)\right]|P\rangle\;,\\
	|(1),(2)\rangle =& -\frac{1}{\sqrt{2}}\left[\left(\frac{w_{-1}}{2}+c_{-1}\right) \left(\chi_{-\frac{1}{2}}-\psi_{-\frac{1}{2}}\right)+\right.\\
	&\left.+\left(\frac{1}{2} h_{-1} \left(\chi_{-\frac{1}{2}}-\psi_{-\frac{1}{2}}\right)+\chi_{-\frac{3}{2}}-\psi_{-\frac{3}{2}}\right)\right]|P\rangle\;,\\
	|(1),(1,1)\rangle =& \frac{1}{\sqrt{2}}\left[-\left(\frac{w_{-1}}{2}+c_{-1}\right) \left(\chi_{-\frac{1}{2}}-\psi_{-\frac{1}{2}}\right)+\right.\\
	&\left.+\left(\frac{1}{2} h_{-1} \left(\chi_{-\frac{1}{2}}-\psi_{-\frac{1}{2}}\right)+\chi_{-\frac{3}{2}}-\psi_{-\frac{3}{2}}\right)\right]|P\rangle\;.
\end{aligned}
\ee
Since the equations above for the vectors labeled by pairs with two non-empty diagrams are written only for the specific value of central charge $\hat{c}=1$ (which correspondence to $b=i$) the total matrix of the scalar products is calculated for this special value:
\be
\langle \vec{\lambda}^1 | \vec{\mu}^1 \rangle = \mathrm{diag}(4,4,4,4,4,4,4,4)\;.
\ee
It is consistent with the AGT relation~\eqref{AGTrel} only is the normalization functions in~\eqref{OldNewOrthBasis} are fixed as follows:
\be
\ba{l}\label{Omega3}
\Omega_{((2),(1))}(P)=\pm 2P(2i-2P)\;,\\
\Omega_{((1,1),(1))}(P)=\pm 2P(2i+2P)\;,\\
\Omega_{((1),(2))}(P)=\pm 2P(2i+2P)\;,\\
\Omega_{((1),(1,1))}(P)=\pm 2P(2i-2P)\;.
\ea
\ee
For the computations we fix $\Omega_{((2),(1))}(P)=2P(2i-2P)$, $\Omega_{((1,1),(1))}(P)=-2P(2i+2P)$ and $\Omega_{\lambda_1,\lambda_2}(P)=\Omega_{\lambda_2,\lambda_1}(-P)$.
\paragraph{Level 2:}
The special basis at this level consists of 16 vectors. 10 of them are labeled by pairs of diagrams with one diagram in the pair being empty:
\be
\ba{l}
\mathbf{|(4),\varnothing\rangle}=J_{(4)}^{(2)}(a_k^{(2)})|P\rangle=\\
=\left[-\left(\frac{1}{2}w_{-1}-c_{-1}\right)^2+b^{-2}\left(\frac{1}{4}h_{-1}^2+h_{-1}\psi_{-1/2}\chi_{-1/2}-\chi_{-3/2}\chi_{-1/2}-\psi_{-3/2}\psi_{-1/2}\right)-\right.\\
\left.-2b^{-2}\left(\frac{1}{2}h_{-2}-\chi_{-3/2}\chi_{-1/2}-\psi_{-3/2}\psi_{-1/2}\right)+2ib^{-1}\left(\frac{1}{2}w_{-1}-c_{-1}\right)\left(\frac{1}{2}h_{-1}+\chi_{-1/2}\psi_{-1/2}\right)-\right.\\
\left.-2ib^{-1}\left(\frac{1}{2}w_{-2}-c_{-2}\right)\right]|P\rangle\;,\\[0.7em]

\mathbf{|(3,1),\varnothing\rangle}=\\
=\frac{1}{4}b^{-2}\left(-4 b^2 c_{-1}^2-b^2 w_{-1}^2+h_{-1}^2+4 \chi_{-3/2}\chi_{-1/2}+4 \psi_{-3/2}\psi_{-1/2}-2 i \left(b^2-1\right) b c_{-1}h_{-1}-\right.\\
\left.-4 i \left(b^2-1\right) b c_{-1}\psi_{-1/2}\chi_{-1/2}+4 b^2 c_{-1}w_{-1}+i \left(b^2-1\right) b w_{-1}h_{-1}-2 \left(b^2-1\right) h_{-2}+\right.\\
\left.+2 i \left(b^2-1\right) b w_{-1}\psi_{-1/2}\chi_{-1/2}-4 \left(b^2-1\right) \chi_{-3/2}\psi_{-1/2}-4 \left(b^2-1\right) \psi_{-3/2}\chi_{-1/2}+\right.\\
\left.+8 i b c_{-2}-4 i b w_{-2}+4 h_{-1}\psi_{-1/2}\chi_{-1/2}\right)|P\rangle\;,\\[0.7em]

\mathbf{|(2,2),\varnothing\rangle}=\left(-c_{-1}^2+\frac{1}{4} h_{-1}^2-\frac{1}{4} w_{-1}^2+\chi_{-3/2}\chi_{-1/2}+\right.\\
\left.+\psi_{-3/2}\psi_{-1/2}+iQc_{-2}-\frac{1}{2}iQw_{-2}+c_{-1}w_{-1}+h_{-1}\psi_{-1/2}\chi_{-1/2}\right)|P\rangle\;,\\[0.7em]

\mathbf{|(2,1,1),\varnothing\rangle}=\\
\frac{1}{4}b^{-1}\left(b^3 h_{-1}^2+4 b^3 \chi_{-3/2}\chi_{-1/2}+4 b^3 \psi_{-3/2}\psi_{-1/2}-4 b c_{-1}^2-b w_{-1}^2+4 b^3 h_{-1}\psi_{-1/2}\chi_{-1/2}+\right.\\
\left.+2 i \left(b^2-1\right) c_{-1}h_{-1}+4 i \left(b^2-1\right) c_{-1}\psi_{-1/2}\chi_{-1/2}+8 i b^2 c_{-2}-i \left(b^2-1\right) w_{-1}h_{-1}+\right.\\
\left.+2 \left(b^2-1\right) b h_{-2}-2 i \left(b^2-1\right) w_{-1}\psi_{-1/2}\chi_{-1/2}+4 \left(b^2-1\right) b \chi_{-3/2}\psi_{-1/2}+\right.\\
\left.+4 \left(b^2-1\right) b \psi_{-3/2}\chi_{-1/2}-4 i b^2 w_{-2}+4 b c_{-1}w_{-1}\right)|P\rangle\;,\\[0.7em]

\mathbf{|(1,1,1,1),\varnothing\rangle}=\\
=\left(\frac{1}{4} \left(b^2 h_{-1}^2+4 b^2 \chi_{-3/2}\chi_{-1/2}+4 b^2 \psi_{-3/2}\psi_{-1/2}-4 c_{-1}^2-w_{-1}^2+4 b^2 h_{-1}\psi_{-1/2}\chi_{-1/2}-\right.\right.\\
\left.\left.-4 b^2 h_{-2}-8 b^2 \chi_{-3/2}\psi_{-1/2}-8 b^2 \psi_{-3/2}\chi_{-1/2}-8 i b c_{-1}\psi_{-1/2}\chi_{-1/2}+8 i b c_{-2}+2 i b w_{-1}h_{-1}+\right.\right.\\
\left.\left.+4 i b w_{-1}\psi_{-1/2}\chi_{-1/2}-4 i b w_{-2}\right)-i b c_{-1}h_{-1}+c_{-1}w_{-1}\right)|P\rangle\;.
\ea
\ee
One can get the remaining 5 expressions taking the screened operators $c_k^\mathrm{R}$ and $\psi_r^\mathrm{R}$ instead of $c_k$ and $\psi_r$ for the corresponding states. There are also 6 states labeled by two non-empty diagrams. Taking $b=i$ one can derive them from  the definition \eqref{C1orthBasisUglov} using the expressions  summed up in the table above:
\be
\begin{aligned}
|(2,1),(1)\rangle&= \frac{1}{2}\left(\chi_{-3/2}+\psi_{-3/2}\right)\left(\chi_{-1/2}-\psi_{-1/2}\right)|P\rangle\;,\\
|(1),(2,1)\rangle&=\frac{1}{2}\left(\chi_{-3/2}-\psi_{-3/2}\right)\left(\chi_{-1/2}+\psi_{-1/2}\right)|P\rangle\;,\\
|(2),(2)\rangle&=\left(c_{-1}^2-\frac{1}{4} h_{-1}^2-\frac{1}{4} w_{-1}^2+\right.\\
 &\left.+\chi_{-3/2}\chi_{-1/2}+\psi_{-3/2}\psi_{-1/2}+2 c_{-1}\psi_{-1/2}\chi_{-1/2}+\frac{1}{2} w_{-1}h_{-1}\right)|P\rangle\;,
\end{aligned}
\ee
\be
\begin{aligned}
|(2),(1,1)\rangle&=\left(c_{-1}^2+\frac{1}{4} h_{-1}^2-\frac{1}{4} w_{-1}^2-\right.\\
&\left.-\chi_{-3/2}\chi_{-1/2}-\psi_{-3/2}\psi_{-1/2}+c_{-1}h_{-1}+w_{-1}\psi_{-1/2}\chi_{-1/2}\right)|P\rangle\;,\\
|(1,1),(2)\rangle&=\left(c_{-1}^2+\frac{1}{4} h_{-1}^2-\frac{1}{4} w_{-1}^2-\right.\\
&\left.-\chi_{-3/2}\chi_{-1/2}-\psi_{-3/2}\psi_{-1/2}-c_{-1}h_{-1}-w_{-1}\psi_{-1/2}\chi_{-1/2}\right)|P\rangle\;,\\
|(1,1),(1,1)\rangle&=\left(c_{-1}^2-\frac{1}{4} h_{-1}^2-\frac{1}{4} w_{-1}^2+\right.\\
&\left.+\chi_{-3/2}\chi_{-1/2}+\psi_{-3/2}\psi_{-1/2}-2 c_{-1}\psi_{-1/2}\chi_{-1/2}-\frac{1}{2} w_{-1}h_{-1}\right)|P\rangle\;.
\end{aligned}
\ee
The matrix of the scalar products for these basis elements is diagonal and the norms of the first 10 states are in agreement with the AGT relation~\eqref{AGTrel}. To satisfy this relation for the last 6 states one should fix the normalization functions \eqref{OldNewOrthBasis} such that:
\be\label{Omega4}
\ba{l}
\Omega^2_{((2,1),(1))}(P)=\Omega^2_{((1),(2,1))}(P)= \left(16 P^2(1+P^2)\right)^2\;,\\[0.5em]
\Omega^2_{((2),(2))}(P)=\Omega^2_{((1,1),(1,1))}(P)= \left(4 (1+P^2)\right)^2\;,\\[0.5em]
\Omega^2_{((2),(1,1))}(P)=\Omega^2_{((1,1),(2))}(P)= \left(4P^2\right)^2\;.
\ea
\ee 

\section{Chain vectors and the special basis}\label{ChainVectorsAndSpecBasis}
In this appendix we define the instanton partition functions \cite{Belavin:2011tb} (see also \cite{Fucito:2004ry}) and derive equations \eqref{ChainSpecBasis1}, \eqref{ChainSpecBasis2}.

\paragraph{Instanton partition functions.}
The functions on the right hand side of \eqref{ChainNorms} are
\be
\ba{l}\label{ZZZ}
Z_{\mathrm{f}}^{\mathrm{sym}}\left(\mu_{i}, \vec{P}, \vec{\lambda}^\sigma\right)=\prod_{i=1}^{4} \prod_{\alpha=1}^{2} \prod_{s \in \lambda_{\alpha}^\sigma,s-\mathrm{white}}\left(\phi\left(P_{\alpha}, s\right)+\mu_{i}\right)\;,\\
Z_{\text {vec }}^{\text {sym }}(\vec{P}, \vec{\lambda}^\sigma)=\left(\prod_{\alpha, \beta=1}^{2} \prod_{s \in S(\lambda_\alpha,\lambda_\beta)} E\left(P_{\alpha}-P_{\beta}, \lambda_{\alpha}, \lambda_{\beta} | s\right)\left(Q-E\left(P_{\alpha}-P_{\beta}, \lambda_{\alpha}, \lambda_{\beta} | s\right)\right)\right)^{-1}
\ea
\ee
with
\be
\ba{l}
\phi(P,s)=(i_s-1)b+(j_s-1)b^{-1}+P\;,\\
E\left(P,\lambda,\mu|s\right)=P+b\left(l_\lambda(s)+1\right)-b^{-1}a_\mu(s)\;,
\ea
\ee
where $i_s$ and $j_s$ are the coordinates of the cell $s$ in the north-south and west-east directions correspondingly, such that for the angle cell $i=1$ and $j=1$. In the second equation in~\eqref{ZZZ} the product goes over a set of cells $S(\lambda_{\alpha},\lambda_\beta)$ such that $s\in S(\lambda_\alpha,\lambda_\beta)\Longleftrightarrow s\in \lambda_\alpha$ and $l_{\lambda_\alpha}(s)+a_{\lambda_\beta}(s)+1 = 0~ \mathrm{mod}~2$. 

The bifundamental part of the instanton partition function used in \eqref{AGTrel} is
\begin{equation}\label{AGT}
\begin{aligned}
Z_{\text{bif}}\left(\alpha\left|P',\vec{\mu}^\sigma, P, \vec{\lambda}^{\tilde{\sigma}}\right.\right)=\!\!\!\!
\prod_{S\left(\lambda_{i},\mu_{j} \right)}\!\!\!\left(Q-E(P_i-P_j^\prime,\lambda_i,\mu_j|s)-\alpha\right)
\!\!\!\prod_{S\left(\mu_{j}, \lambda_{i}\right)}\!\!\!\left(E(P_j^\prime -P_i,\mu_j,\lambda_i|s)-\alpha\right)
\end{aligned}
\end{equation}
with $\vec{P}=(P,-P)$. Here the product goes over the sets of the cells $S\left(\lambda^{\tilde{\sigma}}, \mu^{\sigma}\right)$ such that $s \in S\left(\lambda^{\tilde{\sigma}}, \mu^{\sigma}\right) \Longleftrightarrow s \in \lambda,$ and $l_{\lambda}(s)+a_{\mu}(s)+1+\sigma-\tilde{\sigma} \equiv 0 \bmod 2$.

Alternatively, the functions $Z_{\mathrm{f}}^{\mathrm{sym}}$ and $Z_{\mathrm{vec}}^{\mathrm{sym}}$  in~\eqref{ZZZ} can be expressed in terms of the function $ Z_{\text{bif}}$:
\be
\ba{l}
Z_{\text {vec }}^{\text {sym }}(\vec{P}, \vec{\lambda}^\sigma)=Z_{\mathrm{bif}}\left(0|P,\vec{\lambda}^\sigma,P,\vec{\lambda}^\sigma\right)^{-1}\;,\\
Z_{\mathrm{f}}^{\mathrm{sym}}\left(\mu_{i}, \vec{P}, \vec{\lambda}\right)=Z_{\text{bif}}\left(\alpha_1\left|P_2,\vec{\varnothing}^0, P, \vec{\lambda}^\sigma\right.\right)Z_{\text{bif}}\left(\alpha_3\left|P,\vec{\lambda}^\sigma,P_4,\vec{\varnothing}^0\right.\right)\;.
\ea
\ee

\paragraph{Scalar products between the chain vectors and the basis elements.}
The presence of the orthogonal basis in the module of $\mathcal{A}(2,2)$ makes it possible to use the resolution of identity
\be
\mathbb{1}=\sum_{\vec{\lambda}}\frac{|\vec{\lambda}\rangle \langle \vec{\lambda}|}{\langle \vec{\lambda}|\vec{\lambda}\rangle}
\ee
between the chain vectors in the scalar product \eqref{ChainNorms} having as a result the equality of two sums going over the same sets of diagrams
\be
\sum_{\vec{\lambda}}\frac{_{12}\langle N|\vec{\lambda}\rangle \langle \vec{\lambda}|N\rangle_{34}}{\langle \vec{\lambda}|\vec{\lambda}\rangle}=
\begin{cases}
	\sum_{\vec{\lambda}} Z_{\mathrm{f}}^{\mathrm{sym}}\left(\mu_{i}, \vec{P}, \vec{\lambda}\right) Z_{\mathrm{vec}}^{\mathrm{sym}}(\vec{P}, \vec{\lambda})\;,\quad\text{for integer }N\\
	\sum_{\vec{\lambda}} 2 Z_{\mathrm{f}}^{\mathrm{sym}}\left(\mu_{i}, \vec{P}, \vec{\lambda}\right) Z_{\mathrm{vec}}^{\mathrm{sym}}(\vec{P}, \vec{\lambda})\;,\quad\text{for half-integer }N
\end{cases}
\ee
This leads to the equality for the corresponding elements of the sum
\be
\frac{_{12}\langle N|\vec{\lambda}\rangle \langle \vec{\lambda}|N\rangle_{34}}{\langle \vec{\lambda}|\vec{\lambda}\rangle}=
\begin{cases}
	Z_{\mathrm{f}}^{\mathrm{sym}}\left(\mu_{i}, \vec{P}, \vec{\lambda}\right) Z_{\mathrm{vec}}^{\mathrm{sym}}(\vec{P}, \vec{\lambda})\;,\quad\text{for integer }N\\
	2 Z_{\mathrm{f}}^{\mathrm{sym}}\left(\mu_{i}, \vec{P}, \vec{\lambda}\right) Z_{\mathrm{vec}}^{\mathrm{sym}}(\vec{P}, \vec{\lambda})\;,\quad\text{for half-integer }N
\end{cases}
\ee
Using \eqref{AGTrel} and \eqref{OldNewOrthBasis} one can show that
\be
\langle \vec{\lambda}|\vec{\lambda}\rangle=\frac{_{\vec{\lambda}^\sigma}\langle P | P\rangle_{\vec{\lambda}^\sigma}}{\Omega_{(\lambda_1,\lambda_2)}^2(P)}=\frac{Z_{\mathrm{bif}}\left(0|P,\vec{\lambda}^\sigma,P,\vec{\lambda}^\sigma\right)}{\Omega_{(\lambda_1,\lambda_2)}^2(P)}=\frac{1}{\Omega_{(\lambda_1,\lambda_2)}^2(P) Z^{\mathrm{sym}}_{\mathrm{vec}}(\vec{P},\vec{\lambda})}\;.
\ee
Therefore one can suggest
\be
\Omega_{(\lambda_1,\lambda_2)}^2(P)~_{12}\langle N|\vec{\lambda}\rangle \langle \vec{\lambda}|N\rangle_{34}=
\begin{cases}
	Z_{\mathrm{f}}^{\mathrm{sym}}\left(\mu_{i}, \vec{P}, \vec{\lambda}\right),\quad\text{for integer }N\\
	2 Z_{\mathrm{f}}^{\mathrm{sym}}\left(\mu_{i}, \vec{P}, \vec{\lambda}\right) ,\quad\text{for half-integer }N
\end{cases}
\ee
where the parameters $\mu_i$ are defined in \eqref{mu}.
Equating the terms with the same dependence on $\mu_i$ we get the equations \eqref{ChainSpecBasis1} and \eqref{ChainSpecBasis2}.

\section{Level 2 computations} \label{Level2}
For this level the chain vector is
\be
\ba{l}
|2\rangle_{3,4} = |2\rangle^{SV}_{3,4}+|1\rangle^{SV}_{3,4}~|1\rangle^{H\oplus\widehat{\text{sl}}(2)_2}_{3,4}+|2\rangle^{H\oplus\widehat{\text{sl}}(2)_2}_{3,4}=\\
=\left(\beta_2^{SV}L_{-2}+\beta_{1,1}^{SV}L_{-1}^2+\beta_{3/2,1/2}^{SV}G_{-3/2}G_{-1/2}\right)|P\rangle+\beta_1^{SV}\beta_1^{H\oplus\widehat{\text{sl}}(2)_2}L_{-1}w_{-1}|P\rangle+\\
+\left(\beta_{1,1}^{H\oplus\widehat{\text{sl}}(2)_2}w_{-1}^2+\beta_2^{H\oplus\widehat{\text{sl}}(2)_2}w_{-2}\right)|P\rangle\;,
\ea
\ee
Using \eqref{chain} and \eqref{HSl22ChainCommRel} one can get the coefficients $\beta$:
\be
\ba{l}
\beta_{1,1}^{H\oplus\widehat{\text{sl}}(2)_2}=-\frac{(Q-\alpha_3)^2}{8}\;, \quad \beta_2^{H\oplus\widehat{\text{sl}}(2)_2}=\frac{i(Q-\alpha_3)}{4}\;,\\
\beta_2^{SV}=\frac{-2 \Delta _3^2 (3 \hat{c}+6 \Delta -2)+2 \Delta _3 \left(2 \hat{c} \Delta +2 \Delta _4 (3 \hat{c}+6 \Delta -2)+\hat{c}+4 \Delta  (\Delta +1)-1\right)+2 \left(\Delta -\Delta _4\right) \left(\hat{c} (\Delta -1)+\Delta _4 (3 \hat{c}+6 \Delta -2)+2 (\Delta -3) \Delta +1\right)}{(3 \hat{c}+16 \Delta -3) \left(2 (\hat{c}-3) \Delta +\hat{c}+4 \Delta ^2\right)}\;,\\
\beta_{1,1}^{SV}=\left[\left(\Delta -\Delta _4\right) \left[3 \hat{c}^2 (\Delta +1)+\hat{c} \left(22 \Delta ^2+\Delta -3\right)-\Delta _4 \left(22 \hat{c} \Delta +3 \hat{c} (\hat{c}+1)+32 \Delta ^2-34 \Delta \right)+\right.\right.\\
\left.+2 \Delta  (\Delta  (16 \Delta -25)+5)\right]+\Delta _3^2 \left(22 \hat{c} \Delta	+3 \hat{c} (\hat{c}+1)+32 \Delta ^2-34 \Delta \right)+\Delta _3 \left[44 (\hat{c}-3) \Delta ^2- \right.\\
\left.\left.-2 \Delta _4 \left(22 \hat{c} \Delta +3 \hat{c} (\hat{c}+1)+32 \Delta ^2-34 \Delta \right)+2 (\hat{c} (3 \hat{c}-10)+13) \Delta +3 (\hat{c}-1) \hat{c}+64 \Delta ^3\right]\right]\times\\
\times\left[4 \Delta  (3 \hat{c}+16 \Delta -3) \left(2 (\hat{c}-3) \Delta +\hat{c}+4 \Delta ^2\right)\right]^{-1}\;,\\
\beta_{3/2,1/2}^{SV}=\frac{2 \Delta _3 \left(\Delta _4 (3 \hat{c}-14 \Delta )-10 \Delta ^2+\Delta \right)+\Delta _3^2 (14 \Delta -3 \hat{c})+\left(\Delta -\Delta _4\right) \left(\Delta  (3 \hat{c}+6 \Delta -2)+\Delta _4 (3 \hat{c}-14 \Delta )\right)}{2 \Delta  (3 \hat{c}+16 \Delta -3) \left(2 (\hat{c}-3) \Delta +\hat{c}+4 \Delta ^2\right)}\;.
\ea
\ee

The scalar products of the special basis labeled by the pairs of diagrams with one of them being empty coincide with  \eqref{ChainSpecBasis1}:
\be
\ba{l}
\langle (4),\varnothing|2\rangle=\frac{(2 P-2 P_3-2 P_4+Q) (2 P-2 P_3+2 P_4+Q) (-4 b+2 P-2 P_3-2 P_4+5 Q) (-4 b+2 P-2 P_3+2 P_4+5 Q)}{16 (2 P+Q) (-2 b+2 P+3 Q)}\;,\\
\langle (3,1),\varnothing|2\rangle=\frac{(2 P-2 P_3-2 P_4+Q) (2 P-2 P_3+2 P_4+Q) (-4 b+2 P-2 P_3-2 P_4+5 Q) (-4 b+2 P-2 P_3+2 P_4+5 Q)}{16 (2 P+Q) (-2 b+2 P+3 Q)}\;,\\
\langle (2,2),\varnothing|2\rangle=\frac{(2 P-2 P_3-2 P_4+Q) (2 P-2 P_3+2 P_4+Q) (2 P-2 P_3-2 P_4+3 Q) (2 P-2 P_3+2 P_4+3 Q)}{32 (P+Q) (2 P+Q)}\;,\\
\langle (2,1,1),\varnothing|2\rangle=\frac{(2 P-2 P_3-2 P_4+Q) (2 P-2 P_3+2 P_4+Q) (4 b+2 P-2 P_3-2 P_4+Q) (4 b+2 P-2 P_3+2 P_4+Q)}{16 (2 P+Q) (2 b+2 P+Q)}\;,\\
\langle (1,1,1,1),\varnothing|2\rangle=\frac{(2 P-2 P_3-2 P_4+Q) (2 P-2 P_3+2 P_4+Q) (4 b+2 P-2 P_3-2 P_4+Q) (4 b+2 P-2 P_3+2 P_4+Q)}{16 (2 P+Q) (2 b+2 P+Q)}\;.
\ea
\ee
The other 5 equations with interchanged first and second diagrams in each pair ($\lambda_1\leftrightarrow\lambda_2$) differs only by the change of the sign of $P$. 
The scalar products for the remaining 6 states are:
\be
\ba{l}
\langle (2,1),(1)|2\rangle=\langle (1),(2,1)|2\rangle=-\frac{(P-P_3-P_4) (P+P_3-P_4) (P-P_3+P_4) (P+P_3+P_4)}{16 P^2 \left(P^2+1\right)}\;,\\
\langle (2),(2)|2\rangle=\langle (1,1),(1,1)|2\rangle=-\frac{(P-P_3-P_4) (P+P_3-P_4) (P-P_3+P_4) (P+P_3+P_4)}{4 \left(P^2+1\right)}\;,\\
\langle (2),(1,1)|2\rangle=\langle (1,1),(2)|2\rangle=-\frac{(P-P_3-P_4) (P+P_3-P_4) (P-P_3+P_4) (P+P_3+P_4)}{4 P^2}\;.
\ea
\ee
For the last 6 states with both non-empty diagrams scalar products with the chain vectors agree with \eqref{ChainSpecBasis1} if the signs for the normalization functions \eqref{Omega4} are fixed such that
\be
\ba{l}
\Omega_{((2,1),(1))}(P)=\Omega_{((1),(2,1))}(P)=- 16 P^2(1+P^2)\;,\\
\Omega_{((2),(2))}(P)=\Omega_{((1,1),(1,1))}(P)=- 4 (1+P^2)\;,\\
\Omega_{((2),(1,1))}(P)=\Omega_{((1,1),(2))}(P)=- 4P^2\;.
\ea
\ee

\end{document}